\documentclass[a4paper,11pt]{article}
\pdfoutput=1 % if your are submitting a pdflatex (i.e. if you have
\usepackage{jheppub} % for details on the use of the package, please
\usepackage{braket}
\usepackage{float}
\usepackage{cancel}
\usepackage[retainorgcmds]{IEEEtrantools}

\title{\boldmath Quadratic Contributions of Softly Broken Supersymmetry in the Light of Loop Regularization}

\author[a,c,1]{Dong Bai}
\note{On the leave for Department of Physics and Key Laboratory of Modern Acoustics, Nanjing University, Nanjing 210093, China.}
\author[a,b,c]{and Yue-Liang Wu}
\affiliation[a]{Key Laboratory of Theoretical Physics, Institute of Theoretical Physics,\\Chinese Academy of Sciences, Beijing 100190, China}
\affiliation[b]{International Centre for Theoretical Physics Asia-Pasific(ICTP-AP), Beijing 100190, China}
\affiliation[c]{School of Physical Sciences, University of Chinese Academy of Sciences,\\No.19A Yuquan Road, Beijing 100049, China}

\emailAdd{dbai@itp.ac.cn}
\emailAdd{ylwu@itp.ac.cn}
\abstract{Loop regularization (LORE) is a novel regularization scheme in modern quantum field theories. It makes no change to the spacetime structure and respects both gauge symmetries and supersymmetry. As a result, LORE should be useful in calculating loop corrections in supersymmetry phenomenology. To demonstrate further its power, in this article we revisit in the light of LORE the old issue of the absence of quadratic contributions (quadratic divergences) in softly broken supersymmetric field theories. It is shown explicitly by Feynman diagrammatic calculations that up to two loops the Wess-Zumino model with soft supersymmetry breaking terms (WZ' model), one of the simplest models with the explicit supersymmetry breaking, is free of quadratic contributions. All the quadratic contributions cancel with each other perfectly, which is consistent with results dictated by the supergraph techniques. }

\begin{document} 
\maketitle
\flushbottom
\section{Introduction}
\label{Introduction}
With the advent of the Higgs boson with a mass of approximately 125 GeV at LHC \cite{Aad:2012tfa,Chatrchyan:2012xdj}, the last missing piece of the Standard Model (SM) has been found. After hard workings for more than fifty years, eventually we have gotten a mathematically consistent theory at hand, which provides unprecedented agreements with numerous experiments up to the TeV scale.
 
In SM, the mass of the Higgs is a free parameter, and at the quantum level it receives large contributions from the ultraviolet (UV) physics at some UV scale (say, $M_c$), due to the presence of quadratic contributions in the Higgs self-energy diagrams. In this article, we prefer to use the concept ``UV contribution" to refer to what is traditionally called  ``UV divergence'', as the former is more compatible with the modern effective-field-theory approach to quantum field theories suggested by K.~G.~Wilson, in which all quantum field theories are defined at some physical UV scale, and the infrared (IR) theories could be obtained from the UV theories by doing renormalization-group transformations \cite{Wilson:1973jj}.

At the one-loop level \cite{Veltman:1980mj}, the effective Higgs mass parameter $m_H^2(M_c/\mu)$ at the low-energy scale $\mu$ is related to the UV parameter $m_H^2(M_c)$ at the UV scale $M_c$ by
 \begin{align}
&m_H^2(M_c/\mu)=m_H^2(M_c)-\frac{6}{(4\pi)^2}\left(y^2_t-\frac{1}{4}\lambda_H-\frac{1}{8}g^2_1-\frac{3}{8}g^2_2\right)\left(M_c^2-\mu^2\right)\nonumber\\
&+\mathrm{\ logarithmic\ contributions},
 \end{align}
 where $y_t$, $g_1$, $g_2$, $\lambda_H$ are the top-quark Yukawa coupling, the $U(1)_\text{Y}$ gauge coupling, the $SU(2)_{\text{L}}$ gauge coupling and the Higgs quartic coupling, respectively. Here we have ignored the contributions from the rest particles in SM, since their couplings to the Higgs boson are much weaker. It has been shown recently in Ref.~\cite{Bai:2014lea} that, quadratic contributions from the SM Higgs sector can induce spontaneous electroweak symmetry breaking at $\Lambda_{\text{EW}}\simeq 750$ GeV, { given the SM parameters measured at the low energy as boundary conditions for the renormalization group equations.} Such a mechanism is dubbed as quantum electroweak symmetry breaking, as quadratic contributions that play a significant role come from quantum loop effects. The symmetry breaking scale $\Lambda_{\text{EW}}\simeq 750$ GeV could then be treated as another fundamental scale of SM besides the electroweak scale $v=246$ GeV.

{Although SM is a unprecedented triumph of human intelligence, it is generally believed that SM is certainly not the last words we can say about nature. And right now, the most urgent question that confronts us is: \emph{What is the characteristic scale for new physics?} In literature, this is often referred to as the \emph{gauge hierarchy problem}. At present, one of the leading candidates for new physics beyond SM is supersymmetry (SUSY). SUSY introduces bosonic/fermionic partners for each SM fermionic/bosonic particle, and puts stringent constraints on their properties. To describe the real nature, SUSY has to be broken at the low energy. In the high-energy phenomenological studies, usually this is achieved by introducing soft-SUSY-breaking terms into the supersymmetric Lagrangian by hand. The state-of-art constraints on SUSY in the real world could be found in Ref.~\cite{Olive:2016xmw}. Besides the doubling of particle species, SUSY and softly broken SUSY are also characterized by other novel properties, among which the most important one is the absence of quadratic contributions. Traditionally, this issue is handled by the supergraph technique \cite{Girardello:1981wz}. Although elegant and powerful, supergraph techniques are quite baroque and less useful in phenomenological studies, where, instead, the traditional Feynman diagrammatic calculations are more relevant, and dimensional regularization (DREG) \cite{tHooft:1972tcz} and dimensional reduction (DRED) \cite{Siegel:1979wq} are usually adopted to redefine UV divergent Feynman integrals.\footnote{Noticeably, it is shown in Ref.~\cite{Martin:1993yx}, DREG and DRED can lead to different running couplings in softly broken SUSY, due to the fact the latter preserves SUSY, while the former doesn't.} However, it is well-known that DREG and DRED cannot track quadratic contributions in the $4-\epsilon$ dimension, which makes them less convenient in studying theoretical aspects of softly broken SUSY such as the aforementioned absence of quadratic contributions.}

{In this article, we would like to convince the readers that loop regularization (LORE) proposed in Ref.~\cite{Wu:2002xa,Wu:2003dd} is an ideal regularization scheme in studying both theoretical and phenomenological properties of SUSY and softly broken SUSY. LORE is believed to be able to preserve various symmetries, including Poincare symmetry, gauge symmetry, SUSY, etc,\footnote{These symmetry-preserving properties of LORE are believed to hold at arbitrary loops. At present, the preservation of Non-Abelian gauge symmetries and SUSY has been verified at one loop by checking the Slavnov-Taylor identities \cite{Cui:2008uv} and SUSY Ward identities \cite{Cui:2008bk}, while the preservation of Abelian gauge symmetries has been verified at two loops by checking the Ward identities \cite{Huang:2012iu}.\label{Preservation}} and has already been applied in several studies, such as the one-loop renormalization of Non-Abelian gauge theories \cite{Cui:2008uv}, the study of composite Higgs model \cite{Dai:2003ip}, the gravitational corrections to the running of gauge couplings \cite{Tang:2008ah,Tang:2010cr,Tang:2011gz}, the renormalization of supersymmetric field theories \cite{Cui:2008bk}, the trace anomaly in quantum electrodynamics (QED) \cite{Cui:2011za}, the diphoton channel of the Higgs decay \cite{Huang:2011yf}, the quadratic running of the effective Higgs mass parameter. In Ref.~\cite{Huang:2011xh,Huang:2012iu}, LORE has been used to calculate two-loop quantum corrections of the $\lambda\phi^4$ theory and QED. In a recent review article \cite{Wu:2013iga}, one of the authors (YLW) makes a comprehensive review of the underlying philosophy and application scenarios of LORE. Noticeably, LORE provides not only useful tools for physicists to study quantum field theories, but also new challenges for mathematicians. Recently, Ref.~\cite{Chapling:2016kpi,Chapling:2016sfj} prove the three conjectures concerning the asymptotics of sums of products of binomials, powers and logarithms suggested in Ref.~\cite{Wu:2002xa,Wu:2003dd}, and propose closed-form expressions for Irreducible Loop Integrals (ILIs), which are building blocks of LORE. These studies show that LORE is applicable in arbitrary spacetime dimensions, which makes it suitable in studying quantum loop effects of gravitational gauge field theory in six-dimensional spacetime \cite{Wu:2017xzu} and the unified field theory of basic forces and elementary particles with gravitational origin of gauge symmetry in nineteen-dimensional hyper-spacetime\cite{Wu:2017rmd}.  To demonstrate the power of LORE, we calculate in the later parts of this article the two-loop quadratic contributions of WZ' model, i.e., Wess-Zumino model with soft SUSY breaking terms, and show explicitly the cancellation of all quadratic contributions.}

The rest parts of this article are organized as follows: In Section 2, we give a practical introduction to LORE. In Section 3, we use LORE to calculate quadratic contributions in WZ' model, up to two loops. In Section 4, we conclude with some final remarks. We also include several Appendices to provide some technical details. Besides the aforementioned motivation to provide new tools to study quadratic contributions in SUSY phenomenology, the results of this article could be interpreted further as follows. As mentioned in Footnote \ref{Preservation}, the SUSY preservation of LORE has been verified at one loop by checking Slavnov-Taylor identities directly \cite{Cui:2008bk}. It is desirable to verify explicitly the applicability of LORE to SUSY at two loops and beyond. As complete two-loop calculations are quite complicated, it is wise to first verify some important features of SUSY and softly broken SUSY, such as the absence of quadratic contributions, and approach the final goal step by step. 

\section{A Practical Guide to LORE}
\label{LORE}
In this section, we shall give a practical introduction to LORE. The viewpoint adopted here is slightly different from Ref.~\cite{Wu:2002xa,Wu:2003dd}. And we mainly concentrate on how to use LORE to do the realistic calculations in phenomenology. Readers who want to know more about LORE are recommended to go to Ref.~\cite{Wu:2013iga} for a more comprehensive introduction. 

%%%%%%%%%%%%%%%%%%%%%%%%%%%%%%% SUBSECTION 2.1 %%%%%%%%%%%%%%%%%%%%%%%%%%%%%%%%
\subsection{General Features}
In this subsection, we would like to provide a comparison between LORE and other regularization schemes in literature, making clear the differences between them. The common regularization schemes include sharp cut-off regularization, Pauli-Villars regularization \cite{Pauli:1949zm}, higher derivative regularization \cite{Iliopoulos:1974zv},  DREG, DRED, etc.\footnote{For an introduction to recent developments and comparisons of regularization scheme, we recommend Ref.~\cite{Gnendiger:2017pys}, which also contains a discussion on other four-dimensional regularization schemes such as the implicit regularization (IREG) \cite{Battistel:1998sz,BaetaScarpelli:2001ix,BaetaScarpelli:2000zs} and four-dimensional renormalization (FDR) \cite{Pittau:2012zd}. Although not included in our current goals, comparing LORE with these regularization schemes would be interesting and important.  Hopefully, we could handle this issue in future studies.} The philosophy underlying these regularization schemes can be summarized as follows:
\begin{enumerate}
\item First, the Lagrangians of quantum field theories have to be modified in some ways. In the sharp cut-off regularization, the Lagrangians are reformulated on a discrete lattice; in the Pauli-Villars regularization, extra Pauli-Villars ghost fields and interaction vertices are added into the Lagrangian; in the higher derivative regularization, higher derivative interactions are added in; in DREG and DRED, the Lagrangians are reformulated in the $D=4-\epsilon$ spacetime, rather than the ordinary $D=4$ Minkowski spacetime, and one needs to introduce the extra fields called $\epsilon$-scalars when using DRED to regularize gauge field theories.  

\item The Feynman rules of the Lagrangian then have to be modified correspondingly. In the sharp cut-off regularization, the integrals over loop momenta are cut off at some energy scale; in Pauli-Villars regularization, extra propagators and vertices of the Pauli-Villars ghosts are introduced; in the higher derivative regularization, no extra fields are introduced, but the propagators of the existing fields are modified; in the DREG and DRED, the dimension of the momentum integrals are changed from 4 to $4-\epsilon$, and new propagators and interaction vertices are needed where the $\epsilon$-scalars mentioned above appear. Also, when using DREG and DRED to handle models containing chiral fermions, one needs extra rules to manipulate the Levi-Civita symbol $\epsilon_{\mu\nu\rho\sigma}$ and $\gamma^5$, which can lead to mathematical inconsistencies \cite{Siegel:1980qs}.

\item Finally, the Feynman integrals corresponding to the Feynman diagrams have to be reformulated using the new Feynman rules derived above. For the regularization schemes mentioned above, at this stage, all the Feynman integrals become finite and thus mathematically well-defined.
\end{enumerate}   

\begin{figure}
\centering
\includegraphics[scale=0.35]{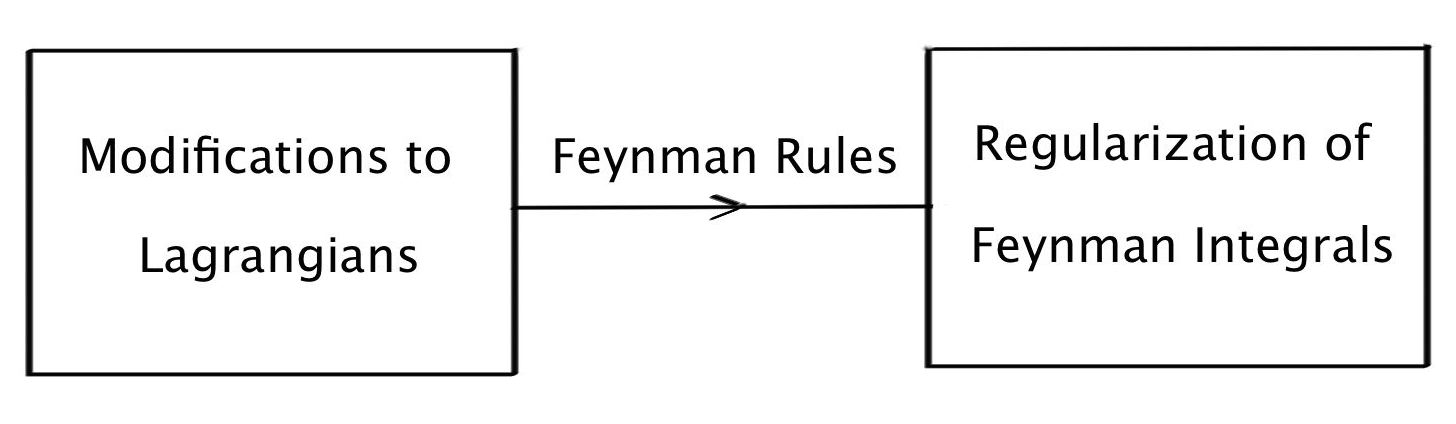}
\caption{Mediation Scenario for the Regularization of Feynman Integrals}
\label{MediationScenarioOfRegularization}
\end{figure}

The above steps are summarized diagrammatically in Fig.~\ref{MediationScenarioOfRegularization}. In the following, we shall call this way to construct regularization schemes the \emph{mediation scenario}, just to emphasize the role played by Feynman rules in transmitting the modifications to the Lagrangians into the target Feynman integrals, and thereby make them mathematically well-defined.
%
%%%%%%%%%%%%%%%%%%%%%%%%%%%%%% FIGURE 1 %%%%%%%%%%%%%%%%%%%%%%%%%%%%%%%%%%%%%
%%%%%%%%%%%%%%%%%%%%%%%%%%%%%%%%%%%%%%%%%%%%%%%%%%%%%%%%%%%%%%%%%%%%%%%%%

LORE is different from all the aforementioned regularization schemes in the sense that its underlying philosophy is different. Instead of utilizing Feynman rules as messengers, it redefines the divergent Feynman integrals directly. So, when using LORE we do not need to modify either the Lagrangians or the Feynman rules directly. We don't need any unphysical fields, any extra vertices, any modifications to the propagators of the existing fields, or any departures from the ordinary spacetime structures. The essential reason for the fact that LORE preserves SUSY is that SUSY can be regarded as a spacetime symmetry as it extends the usual Minkowski spacetime into the so-called superspace which contains not only the usual commutative coordinates but also anti-commutative coordinates. As emphasized above, unlike the popular DREG and DRED, LORE does not change the spacetime structure at all. So it is physically straightforward to see that LORE should preserve SUSY perfectly.

{For a general divergent Feynman integral, it is shown in Ref.~\cite{Huang:2011xh,Huang:2012iu} that the structures of UV contributions can be extracted in the following way
\begin{equation}
I=I_\text{FP} \otimes I_\text{UVDP} \otimes I_\text{ILI},
\label{LOREstructure}
\end{equation}
by exploiting Bjorken-Drell's analogy between Feynman diagrams and electrical circuits \cite{Bjorken:1965zz}.} Here, $I$ stands for the divergent Feynman integral, $I_\text{FP}$ for integrals over Feynman parameters, $I_\text{UVDP}$ for integrals over the ultraviolet-divergence-preserving (UVDP) parameters introduced in Ref.~\cite{Huang:2011xh,Huang:2012iu}, and $I_\text{ILI}$ for the ILIs. The $\otimes$ operation here is introduced for heuristic reasons, and $I_x \otimes I_y$ roughly means $\int\!\mathrm{d}x I_x\!\int\!\mathrm{d}y I_y$. Generally, $I_y$ depends on both $x$ and $y$.  The following discussions do not rely on the precise definition of the operator $\otimes$. Before giving explicit definitions of these concepts, we want to emphasize some general features of $I_\text{FP}$, $I_\text{UVDP}$, $I_\text{ILI}$ first  \cite{Huang:2011xh,Huang:2012iu}:

\begin{enumerate}
\item Generally, the UV-contribution structures of the Feynman integral $I$ are encoded entirely in $I_\text{UVDP}$ and $I_\text{ILI}$. In other words, the Feynman parameter integrals $I_\text{FP}$ contain no UV contributions.

\item The overall UV contribution of $I$ is solely encoded in $I_\text{ILI}$. If the Feynman integral $I$ has any tensor structure, the tensor structure is also encoded entirely in $I_\text{ILI}$.

\item The UV subcontributions in $I$ are encoded in the UVDP integrals $I_\text{UVDP}$. And it is shown that there is a one-to-one correspondence between UV subcontributions (UV subdivergences) in $I$ and the UV contributions in $I_\text{UVDP}$.
\end{enumerate}

For the one-loop Feynman integrals, there is no UV subcontribution in the Feynman integrals. Thus, in this case we do not need UVDP integrals, and Eq.~\eqref{LOREstructure} can be simplified to
\begin{equation}
I=I_\text{FP} \otimes I_\text{ILI}.
\label{LOREstructure1loop}
\end{equation}

%%%%%%%%%%%%%%%%%%%%%%%%%%%%%% END OF SUBSECTION 2.1 %%%%%%%%%%%%%%%%%%%%%%%%%%%%%%%%
%%%%%%%%%%%%%%%%%%%%%%%%%%%%%%%%%%%%%%%%%%%%%%%%%%%%%%%%%%%%%%%%%%%%%%%%%%%%
%%%%%%%%%%%%%%%%%%%%%%%%%%%%%%% SECTION 2.2 %%%%%%%%%%%%%%%%%%%%%%%%%%%%%%%%%%%%%%
\subsection{ILI and LORE}
Let's start with the definitions of ILIs (i.e.~irreducible loop integrals):
\begin{equation}
I_{-2\alpha}=\int\!\mathrm{d}^4k \frac{1}{(k^2-\mathcal{M}^2)^{2+\alpha}}.
\end{equation}
Here $\alpha=-1, 0, 1, 2,\cdots$, and the number ($-2\alpha$) in the subscript labels the superficial degrees of UV contributions of ILIs. $\mathcal{M}$ is generally a function of the physical masses $m_i$, external momenta $p_i$, and other parameters introduced during the calculation, e.g. Feynman parameters $x_i$. Formally, one has $\mathcal{M}=\mathcal{M}(m_i, p_i, x_i, \cdots)$.

One can also introduce extra tensor structures into ILIs:
\begin{align}
&I_{-2\alpha}^{\mu\nu}=\int\!\mathrm{d}^4k\frac{k^{\mu}k^{\nu}}{(k^2-\mathcal{M}^2)^{3+\alpha}},\\
&I_{-2\alpha}^{\mu\nu\rho\sigma}=\int\!\mathrm{d}^4k\frac{k^{\mu}k^{\nu}k^{\rho}k^{\sigma}}{(k^2-\mathcal{M}^2)^{4+\alpha}},
\end{align}  
etc. Here $\alpha=-1, 0, 1, 2,\cdots$. These tensor type ILIs are common when doing calculations in gauge field theories.

As mentioned in Section 2.2, at the one-loop level, one can always decompose a Feynman integral $I$ into a ``product" of $I_\text{FP}$ and $I_\text{ILI}$. Let's take the following example to see how this kind of decomposition comes into being. 

\underline{Example}: Decompose $I=\int\!\mathrm{d}^4k\frac{1}{k^2-m^2_1}\frac{1}{(k+p)^2-m^2_2}$ into the form $I_\text{FP}\otimes I_\text{ILI}$.

Using the standard Feynman parametrization, one can easily show that 
\begin{align}
&\:\int\!\mathrm{d}^4k\frac{1}{k^2-m^2_1}\frac{1}{(k+p)^2-m^2_2}\nonumber\\ 
=&\int^1_0\!\mathrm{d}x\int\!\mathrm{d}^4k\frac{1}{\{(1-x)(k^2-m^2_1)+x[(k+p)^2-m^2_2]\}^2}\nonumber\\
=&\int^1_0\!\mathrm{d}x\int\!\mathrm{d}^4k\frac{1}{\{(k+xp)^2-[(1-x)m^2_1+xm^2_2-x(1-x)p^2]\}^2}\nonumber\\
=&\underbrace{\int^1_0\!\mathrm{d}x}_{I_\text{FP}}\underbrace{\int\!\mathrm{d}^4k\frac{1}{(k^2-\mathcal{M}^2)^2}}_{I_\text{ILI}}.
\end{align}
In the last step, one uses the variable shift\footnote{As shown in Ref.~\cite{Cui:2008bk}, such kind of momentum shift is perfectly legal in LORE, even though at this stage, one has not introduced any regularization scheme yet. One can show explicitly that the order of doing momentum shift and LORE does not matter at all, and different orders should give the same results. So here we choose to do the momentum shift first and then use LORE to regularize the integrals.} $k\to k-xp$ and $\mathcal{M}^2=(1-x)m^2_1+xm^2_2-x(1-x)p^2$.

In the above, we have introduced the concepts of ILIs and shown explicitly that at the one-loop level, one can decompose Feynman integrals into ``products" of $I_\text{FP}$ and $I_\text{ILI}$. As emphasized in Section 2.1, the overall UV contribution and the tensor structure of Feynman integrals should always be encoded in $I_\text{ILI}$. And in LORE, in order to regularize the divergent Feynman integral, one has to give proper redefinitions to ILIs. These redefinitions could be found in Ref.~\cite{Wu:2002xa,Wu:2003dd}, and are reproduced in the following:
\begin{align*}
&\underline{Prescription\ 1}:\\
&\int\!\mathrm{d}^4k\frac{1}{k^2-\mathcal{M}^2}:=-i\pi^2\left\{M^2_c-(\mathcal{M}^2+\mu^2_s)\left[\ln\frac{M^2_c}{\mathcal{M}^2+\mu^2_s}-\gamma_{\omega}+1+y_2\left(\frac{\mathcal{M}^2+\mu^2_s}{M^2_c}\right)\right]\right\},\\
&\underline{Prescription\ 2}:\\
&\int\!\mathrm{d}^4k\frac{1}{(k^2-\mathcal{M}^2)^{n+1}}:=\frac{1}{n}\frac{\partial}{\partial\mathcal{M}^2}\int\!\mathrm{d}^4k\frac{1}{(k^2-\mathcal{M}^2)^n},\quad \quad \quad(n\geq 1)\\
&\underline{Prescription\ 3}\ (Consistency\ Conditions):\\
&I^{\mu\nu}_{-2\alpha}:=\frac{1}{2(\alpha+2)}g^{\mu\nu}I_{-2\alpha}, \quad\  I^{\mu\nu\rho\sigma}_{-2\alpha}:=\frac{1}{4(\alpha+2)(\alpha+3)}(g^{\mu\nu}g^{\rho\sigma}+g^{\mu\rho}g^{\nu\sigma}+g^{\mu\sigma}g^{\nu\rho})I_{-2\alpha}.\\
&\text{Here $\alpha=-1, 0, 1, 2, \cdots$ For $\alpha=-1\ or\ 0$, one has}\\
&I^{\mu\nu}_2:=\frac{1}{2}g^{\mu\nu}I_2, \qquad \qquad \qquad I^{\mu\nu\rho\sigma}_2:=\frac{1}{8}(g^{\mu\nu}g^{\rho\sigma}+g^{\mu\rho}g^{\nu\sigma}+g^{\mu\sigma}g^{\nu\rho})I_2,\\
&I^{\mu\nu}_0:=\frac{1}{4}g^{\mu\nu}I_0, \qquad \qquad \qquad I^{\mu\nu\rho\sigma}_0:=\frac{1}{24}(g^{\mu\nu}g^{\rho\sigma}+g^{\mu\rho}g^{\nu\sigma}+g^{\mu\sigma}g^{\nu\rho})I_0.
\end{align*}
Here $M_c$ acts as the UV scale, while $\mu_s$ acts as the infrared (IR) scale. For those theories that are free of IR divergences, $\mu_s$ can be safely set to zero. $\gamma_{\omega}$ equals the Euler constant $\gamma_E$. $y_2(x)=\int^x_0\!\mathrm{d}\sigma\frac{1-e^{-\sigma}}{\sigma}+\frac{1}{x}(1-x-e^{-x})$, and it can be easily shown that when $x\to 0$, $y_2(x)\to 0$. In other words, when $M_c\to \infty$, $y_2$ function in Prescription 1 goes to zero. Prescription 3 plays an extremely important role in preserving gauge symmetries in regularization schemes. It is shown in great details in Ref.~\cite{Wu:2002xa,Wu:2003dd} that a regularization scheme can preserve gauge symmetries only if these consistency conditions are satisfied. The sharp cut-off regularization scheme, which is well-known to break gauge symmetries, does not satisfy these consistency conditions, while DREG and DRED do.

By using the above prescriptions, one can derive explicit expressions for ILIs other than $I_2$, and many of the useful results are summarized in Appendix. Before moving on to the multi-loop calculations, we want to emphasize the following two points:
\begin{enumerate}
\item The above treatment of LORE is practically oriented, and we aim to explain how to use LORE to do realistic calculations. Although many results might seem ad hoc for some readers, for instance, we explain neither why $I_2$ should be defined as that in Prescription 1, nor where the non-trivial $y_2$ function comes from, these are actually well-motivated and we recommend Ref.~\cite{Wu:2002xa,Wu:2003dd} for further details.
  
\item In LORE, one can track quadratic contributions along with logarithmic contributions at the same time. This can be seen explicitly in Prescription 1, where the integral on the left-hand side, i.e., $I_2$, is quadratic divergent. This is actually highly nontrivial, when taking into consideration that LORE preserves also gauge symmetries. In the popular DREG and DRED, which preserve gauge symmetries as well, quadratic contributions can only be extracted by tracing pole structures of the Feynman integrals at dimensions lower than 4. Practically, this means that one has to carry out a separate calculation to extract quadratic contributions.
\end{enumerate}

We can also extend the above treatment to multi-loop calculations, although often much more complicated due to the appearance of UV subcontributions. As mentioned in Section 2.1, these subcontributions can be fully captured in the UVDP integrals $I_{UVDP}$, and we recommend Ref.~\cite{Huang:2011xh,Huang:2012iu} for a discussion of what UVDP integrals are and how to extract subcontributions from it. In this articles, we shall continue taking a practical viewpoint without going into complicated details. Here, instead of doing the decomposition shown in Eq.~\eqref{LOREstructure} and regularize the divergent subintegrals one by one, we shall try to give out the final results directly in a way that they can be used repeatedly in practical calculations. The key observation comes from the fact that $I_\text{UVDP}\otimes I_\text{ILI}$ actually comprises the so-called $\alpha\beta\gamma$ integrals introduced by `t Hooft and Veltman \cite{tHooft:1972tcz}. 
\begin{equation}
I_\text{UVDP}\otimes I_\text{ILI}=I_{\alpha\beta\gamma}.
\end{equation}
For the two-loop scalar Feynman integrals, a typical $I_{\alpha\beta\gamma}$ is given by\footnote{In the real calculations, one encounters extra complications from nontrivial numerators other than 1. The discussion here only deals with the simplest case of $I_{\alpha\beta\gamma}$, and the general cases can be treated in a similar way.}
\begin{equation}
I_{\alpha\beta\gamma}=\int\!\mathrm{d}^4l_1\!\int\!\mathrm{d}^4l_2\frac{1}{(l^2_1-m^2_1)^{\alpha}}\frac{1}{(l^2_2-m^2_2)^{\beta}}\frac{1}{[(l_1+l_2+p)^2-m^2_3]^{\gamma}} .
\end{equation}
So, instead of Eq.~\eqref{LOREstructure}, one has
\begin{equation}
I=I_\text{FP}\otimes I_{\alpha\beta\gamma}.
\end{equation}
Ref.~\cite{Huang:2011xh,Huang:2012iu} contain comprehensive discussions on calculating the $\alpha\beta\gamma$ integrals with the help of the Bjorken-Drell's electrical-circuit analogy, and show explicitly the one-to-one correspondence between subcontributions in the original $\alpha\beta\gamma$ integrals and those in the UVDP integrals.

The above equation just says that given a general two-loop Feynman integral, we can always using the standard Feynman parametrization to reduce it to the form of a ``product" of Feynman parameter integral $I_\text{FP}$ and $I_{\alpha\beta\gamma}$. Since $I_\text{FP}$ does not contain any UV contributions, to regularize $I$, one just needs to regularize the $I_{\alpha\beta\gamma}$ parts. So for practical purposes, instead of doing the hard work of calculating $I_\text{UVDP}$ and $I_\text{ILI}$ case by case, one just needs the regularized results of $I_{\alpha\beta\gamma}$.  

For the cases $\alpha$, $\beta$ or $\gamma$ equals zero, $I_{\alpha\beta\gamma}$ can be decomposed into two one-loop $I_\text{ILI}$, and to get the regularized results, all one has to do is to use the one-loop results twice. For $\alpha$, $\beta$, $\gamma$$\ \neq 0$, the only case that is relevant to our calculations of quadratic contributions in WZ' model in the next section is $I_{111}$ whose quadratic contributions are given by:
\begin{equation}
I_{111}=\pi^4M^2_c[3(\ln\frac{M^2_c}{q^2_0}-\gamma_{\omega})+1]+\cdots.
\end{equation}
Here, $q_0$ is an arbitrary mass scale introduced to balance the dimension. The rest cases (e.g.~$I_{112}$) just do not contain any quadratic contributions by naive power counting. One can find more discussions about $I_{\alpha\beta\gamma}$ in the Appendix.

Now we have accumulated sufficient information about LORE to finish our calculations. Let's move on to discuss the WZ' model and try to calculate the possible quadratic contributions up to two loops.

\section{Quadratic Contributions in WZ' Model}

WZ' model, i.e., Wess-Zumino model \cite{Wess:1973kz,Wess:1974tw} with soft SUSY breaking terms, is the simplest model for the softly broken SUSY, and is an insightful toy model that shares many important properties of MSSM. The treatment here can be extended straightforwardly to the more complicated models that include gauge bosons since LORE preserves the gauge symmetries as well. The lagrangian of WZ' model is given by
\begin{multline}
\mathcal{L}_{WZ'}=\frac{1}{2}(\partial_{\mu}A)^2-\frac{1}{2}m^2_AA^2+\frac{1}{2}(\partial_{\mu}B)^2-\frac{1}{2}m^2_BB^2+\frac{1}{2}\bar{\psi}(i\cancel{\partial}-m_{\psi})\psi\\
-(m_{\psi}g-\lambda)A^3-(m_{\psi}g+3\lambda)AB^2-g(\bar{\psi}\psi A+i\bar{\psi}\gamma^5\psi B)-\frac{1}{2}g^2(A^2+B^2)^2.
\label{WZ'Lagrangian}
\end{multline} 
In this article, we adopt the $(+,-,-,-)$ convention and $\gamma^5=i\gamma^0\gamma^1\gamma^2\gamma^3$. A and B are real scalar fields, and $\psi$ is a Majorana fermion. g is a dimensionless real coupling and $\lambda$ is the dimension-one soft SUSY breaking parameter. In Eq.~\eqref{WZ'Lagrangian} we have integrated out the so-called auxiliary fields, which is a common practice in phenomenology. The conventions of Feynman rules we adopt in the following calculations are quite standard and can be found in Ref.~\cite{Peskin:1995ev} and Ref.~\cite{Denner:1992vza,Denner:1992me}. 
%%%%%%%%%%%%%%%%%%%%%%%%%%%%%%%%%% TABLE 1 %%%%%%%%%%%%%%%%%%%%%%%%%%%%%%%%%%%%%%%%%%
\begin{table}[t]
\caption{One-loop Counterterms for WZ' Model}
\label{OneLoopCounterterms}
\centering
\begin{tabular}{c c}
\hline
\hline
Counterterm & Result\\
\hline
\\ [-2ex]
$\delta Z^{(1)}_A$ & $-\frac{g^2}{4\pi^2}\ln\frac{M^2_c}{\mu^2}$\\ [1ex]
$\delta Z^{(1)}_{AA}$ & $\left[\frac{3g^2}{8\pi^2}+\frac{g^2m^2_B}{8\pi^2m^2_A}-\frac{3g^2m^2_{\psi}}{2\pi^2m^2_A}+\frac{9(-\lambda+gm_{\psi})^2}{8\pi^2m^2_A}+\frac{(3\lambda+gm_{\psi})^2}{8\pi^2m^2_A}\right]\ln\frac{M^2_c}{\mu^2}$\\ [1ex]
$\delta Z^{(1)}_B$ & $-\frac{g^2}{4\pi^2}\ln\frac{M^2_c}{\mu^2}$\\ [1ex]
$\delta Z^{(1)}_{BB}$ & $[\frac{3g^2}{8\pi^2}+\frac{g^2m^2_A}{8\pi^2m^2_B}-\frac{g^2m^2_{\psi}}{2\pi^2m^2_B}+\frac{(3\lambda+gm_{\psi})^2}{4\pi^2m^2_B}]\ln\frac{M^2_c}{\mu^2}$\\ [1ex]
$\delta Z^{(1)}_{\psi}$ & $-\frac{g^2}{4\pi^2}\ln\frac{M^2_c}{\mu^2}$\\ [1ex]
$\delta Z^{(1)}_{\bar{\psi}\psi}$ & $0$\\ [1ex]
$\delta Z^{(1)}_{AAA}$ & $(\frac{g^2}{4\pi^2}-\frac{3g\lambda}{4\pi^2m_{\psi}})\ln\frac{M^2_c}{\mu^2}$\\ [1ex]
$\delta Z^{(1)}_{ABB}$ & $(\frac{g^2}{4\pi^2}+\frac{9g\lambda}{4\pi^2m_{\psi}})\ln\frac{M^2_c}{\mu^2}$\\ [1ex]
$\delta Z^{(1)}_{\bar{\psi}\psi A}$ & $0$\\ [1ex]
$\delta Z^{(1)}_{\bar{\psi}\psi B}$ & $0$\\ [1ex]
$\delta Z^{(1)}_{AAAA}$ & $\frac{g^2}{4\pi^2}\ln\frac{M^2_c}{\mu^2}$\\ [1ex]
$\delta Z^{(1)}_{BBBB}$ & $\frac{g^2}{4\pi^2}\ln\frac{M^2_c}{\mu^2}$\\ [1ex]
$\delta Z^{(1)}_{AABB}$ & $\frac{g^2}{4\pi^2}\ln\frac{M^2_c}{\mu^2}$\\ [1ex]
\hline
\end{tabular}
\end{table} 
%%%%%%%%%%%%%%%%%%%%%%%%%%%%%%%%%%%%%%%%%%%%%%%%%%%%%%%%%%%%%%%%%%%%%%%%%%%%%%%%%

In this section, we shall calculate the quadratic contributions in the self-energy diagrams of the scalar particle A and the pseudoscalar particle B up to two loops. In these calculations, we have used the \emph{Mathematica} package \emph{FeynArts} \cite{Hahn:2000kx} to generate the relevant Feynman diagrams.

%%%%%%%%%%%%%%%%%%%%%%%%%%%%%%%%%% SUBSECTION 3.1 %%%%%%%%%%%%%%%%%%%%%%%%%%%%%%%%%%%%%
\subsection{One-loop Calculations\label{1LoopCalculation}}
Beyond the tree level, one needs to introduce extra counterterms in order to make the radiative corrections finite. To calculate radiative corrections at two loops, one has to first figure out counterterms at the one-loop level. At the one-loop level, the counterterm lagrangian is given by
\begin{align}
&\mathcal{L}_{ct}=\frac{1}{2}\delta Z^{(1)}_A(\partial_{\mu}A)^2-\frac{1}{2}\delta Z^{(1)}_{AA}m^2_AA^2+\frac{1}{2}\delta Z^{(1)}_B(\partial_{\mu}B)^2-\frac{1}{2}\delta Z^{(1)}_{BB}m^2_BB^2\nonumber\\
&+\frac{1}{2}\delta Z^{(1)}_{\psi}\bar{\psi}i\cancel{\partial}\psi-\frac{1}{2}\delta Z^{(1)}_{\bar{\psi}\psi}m_{\psi}\bar{\psi}\psi-\delta Z^{(1)}_{AAA}m_{\psi}gA^3-\delta Z^{(1)}_{ABB}m_{\psi}gAB^2-\delta Z^{(1)}_{\bar{\psi}\psi A}g\bar{\psi}\psi A\nonumber\\
&-\delta Z^{(1)}_{\bar{\psi}\psi B}ig\bar{\psi}\gamma^5\psi B-\frac{1}{2}\delta Z^{(1)}_{AAAA}g^2A^4-\frac{1}{2}\delta Z^{(1)}_{BBBB}g^2B^4-\delta Z^{(1)}_{AABB}g^2A^2B^2.
\end{align}
At the one-loop level, there are 114 Feynman diagrams that can contribute to the one-loop counterterms. It would be too messy to draw all of these diagrams here. Instead, we choose to present the final results in Table \ref{OneLoopCounterterms} directly. Here we have used the MS renormalization scheme and the parameter $\mu$ is the so-called renormalization scale. Apparently, there is no quadratic contribution at the one-loop level.

When taking the supersymmetric limit $m_A=m_B=m_{\psi}=m$, $\lambda=0$, one has
\begin{align}
&\delta Z^{(1)}_A=\delta Z^{(1)}_B=\delta Z^{(1)}_{\psi}=-\frac{g^2}{4\pi^2}\ln\frac{M^2_c}{\mu^2},\\
&\delta Z^{(1)}_{AA}=\delta Z^{(1)}_{BB}=\delta Z^{(1)}_{AAA}=\delta Z^{(1)}_{ABB}\nonumber\\
&=\delta Z^{(1)}_{AAAA}=\delta Z^{(1)}_{BBBB}=\delta Z^{(1)}_{AABB}=\frac{g^2}{4\pi^2}\ln\frac{M^2_c}{\mu^2},\\
&\delta Z^{(1)}_{\bar{\psi}\psi}=\delta Z^{(1)}_{\bar{\psi}\psi A}=\delta Z^{(1)}_{\bar{\psi}\psi B}=0,
\end{align}
which are nothing but the celebrated non-renormalization theorem \cite{Wess:1974tw,Wess:1973kz,Cui:2008bk}.

%%%%%%%%%%%%%%%%%%%%%%%%%%%%%%%%% END OF SUBSECTION 3.1 %%%%%%%%%%%%%%%%%%%%%%%%%%%%%%%%%%
%%%%%%%%%%%%%%%%%%%%%%%%%%%%%%%%%%%%%%%%%%%%%%%%%%%%%%%%%%%%%%%%%%%%%%%%%%%%%%%%%
%%%%%%%%%%%%%%%%%%%%%%%%%%%%%%%%%% SECTION 3.2 %%%%%%%%%%%%%%%%%%%%%%%%%%%%%%%%%%%%%%%%
\subsection{Two-loop Calculations}

Now we are ready to calculate quadratic contributions in the self-energy diagrams of the scalar particle A and pseudoscalar particle B in WZ' model at the two-loop level. The relevant Feynman diagrams are given in Fig.~\ref{2LoopAA} and Fig.~\ref{2LoopBB}. Here, we only include diagrams that are quadratic divergent according to power counting. 
%%%%%%%%%%%%%%%%%%%%%%%%%%%%%%%%%% FIGURE 2 %%%%%%%%%%%%%%%%%%%%%%%%%%%%%%%%%%%%%%%%%%
\begin{figure}
\centering
\includegraphics[scale=0.9]{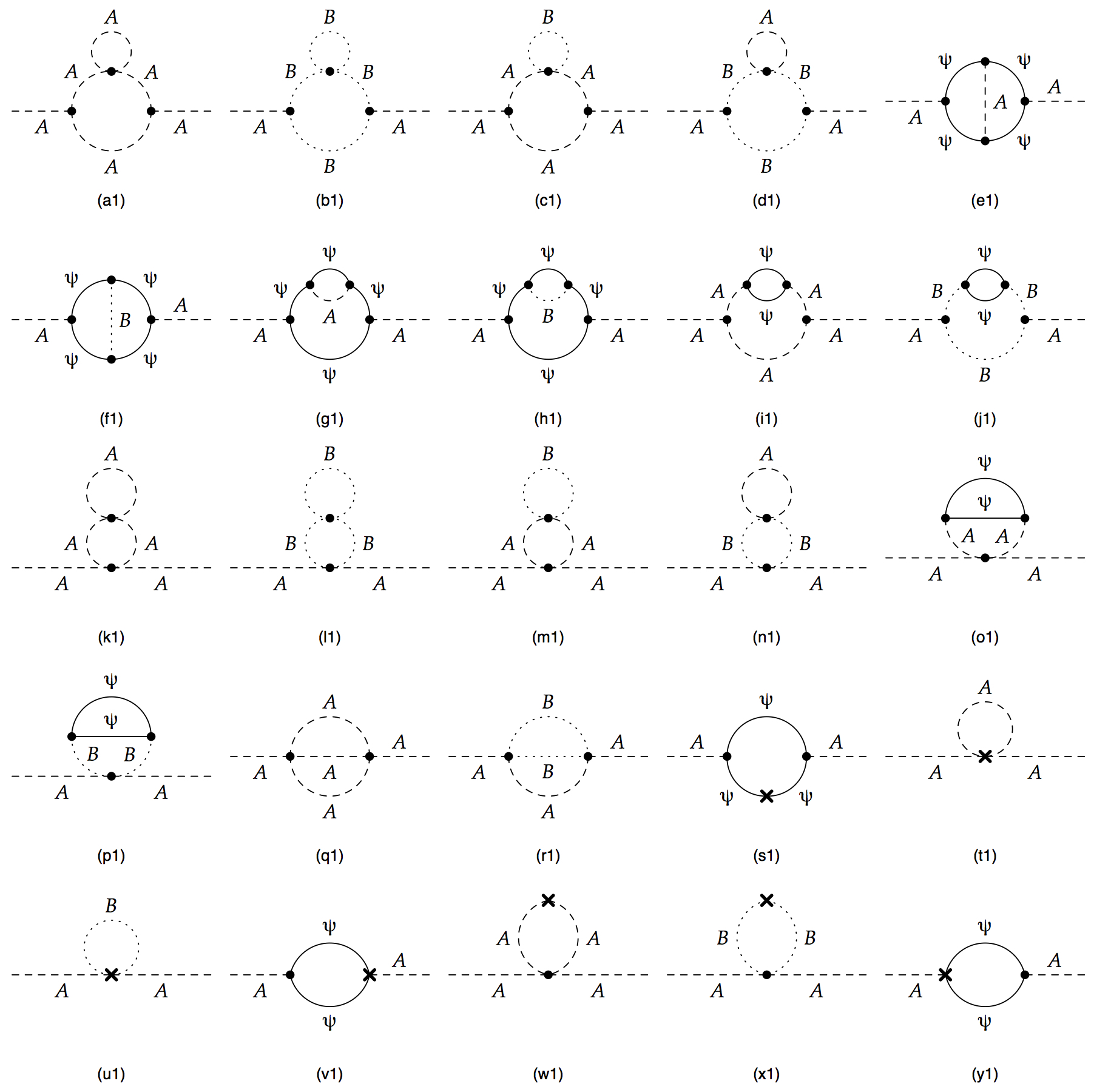}
\caption{Two-loop Self-energy Diagrams for Scalar Particle A}
\label{2LoopAA}
\end{figure}
%%%%%%%%%%%%%%%%%%%%%%%%%%%%%%%%%%%%%%%%%%%%%%%%%%%%%%%%%%%%%%%%%%%%%%%%%%%%%%%%%%

%%%%%%%%%%%%%%%%%%%%%%%%%%%%%%%%%% TABLE 2 %%%%%%%%%%%%%%%%%%%%%%%%%%%%%%%%%%%%%%%%%%%
\begin{table}
\caption{Quadratic Contributions in the Two-Loop Self-energy Diagrams of Scalar Particle A}
\label{2LoopAA}
\centering
\begin{tabular}{p{5cm}c c}
\\[-2ex]
\hline
\hline
Diagram & Quadratic Contributions\\
\hline
\\[-2ex]
\quad(a1) & $-\frac{27g^2(-\lambda+gm_{\psi})^2}{32\pi^4}M^2_c\int^1_0\!\mathrm{d}x\frac{x}{m^2_A-x(1-x)k_1^2}$\\ [1ex]
\quad(b1) & $-\frac{3g^2(3\lambda+gm_{\psi})^2}{32\pi^4}M^2_c\int^1_0\!\mathrm{d}x\frac{x}{m^2_B-x(1-x)k_1^2}$\\ [1ex]
\quad(c1) & $-\frac{9g^2(-\lambda+gm_{\psi})^2}{32\pi^4}M^2_c\int^1_0\!\mathrm{d}x\frac{x}{m^2_A-x(1-x)k_1^2}$\\ [1ex]
\quad(d1) & $-\frac{g^2(3\lambda+gm_{\psi})^2}{32\pi^4}M^2_c\int^1_0\!\mathrm{d}x\frac{x}{m^2_B-x(1-x)k_1^2}$\\ [1ex]
\quad(e1) & $-\frac{g^4}{8\pi^4}M^2_c\left[3\left(\ln\frac{M^2_c}{q^2_0}-\gamma_{\omega}\right)+1\right]$\\ [1ex]
\quad(f1) & $\frac{g^4}{8\pi^4}M^2_c\left[3\left(\ln\frac{M^2_c}{q^2_0}-\gamma_{\omega}\right)+1\right]$\\ [1ex]
\quad(g1) & $-\frac{g^4}{8\pi^4}M^2_c\left[3\left(\ln\frac{M^2_c}{q^2_0}-\gamma_{\omega}\right)+1\right]$\\ [1ex]
\quad(h1) & $-\frac{g^4}{8\pi^4}M^2_c\left[3\left(\ln\frac{M^2_c}{q^2_0}-\gamma_{\omega}\right)+1\right]$\\ [1ex]
\quad(i1) & $\frac{9g^2(-\lambda+gm_{\psi})^2}{8\pi^4}M^2_c\int^1_0\!\mathrm{d}x\frac{x}{m^2_A-x(1-x)k_1^2}$\\ [1ex]
\quad(j1) & $\frac{g^2(3\lambda+gm_{\psi})^2}{8\pi^4}M^2_c\int^1_0\!\mathrm{d}x\frac{x}{m^2_B-x(1-x)k_1^2}$\\ [1ex]
\quad(k1) & $\frac{9g^4}{64\pi^4}M^2_c\left(\ln\frac{M^2_c}{m^2_A}-\gamma_{\omega}\right)$\\ [1ex]
\quad(l1) & $\frac{3g^4}{64\pi^4}M^2_c\left(\ln\frac{M^2_c}{m^2_B}-\gamma_{\omega}\right)$\\ [1ex]
\quad(m1) & $\frac{3g^4}{64\pi^4}M^2_c\left(\ln\frac{M^2_c}{m^2_A}-\gamma_{\omega}\right)$\\ [1ex]
\quad(n1) & $\frac{g^4}{64\pi^4}M^2_c\left(\ln\frac{M^2_c}{m^2_B}-\gamma_{\omega}\right)$\\ [1ex]
\quad(o1) & $\frac{3g^4}{32\pi^4}M^2_c\left(\ln\frac{M^2_c}{q^2_0}+2\ln\frac{m^2_A}{q^2_0}-\gamma_{\omega}+1\right)$\\ [1ex]
\quad(p1) & $\frac{g^4}{32\pi^4}M^2_c\left(\ln\frac{M^2_c}{q^2_0}+2\ln\frac{m^2_B}{q^2_0}-\gamma_{\omega}+1\right)$\\ [1ex]
\quad(q1) & $\frac{3g^4}{32\pi^4}M^2_c\left[3\left(\ln\frac{M^2_c}{q^2_0}-\gamma_{\omega}\right)+1\right]$\\ [1ex]
\quad(r1) & $\frac{g^4}{32\pi^4}M^2_c\left[3\left(\ln\frac{M^2_c}{q^2_0}-\gamma_{\omega}\right)+1\right]$\\ [1ex]
\quad(s1) & $\frac{g^4}{4\pi^4}M^2_c \ln\frac{M^2_c}{\mu^2}$\\ [1ex]
\quad(t1) & $-\frac{3g^4}{32\pi^4}M^2_c \ln\frac{M^2_c}{\mu^2}$\\ [1ex]
\quad(u1) & $-\frac{g^4}{32\pi^4}M^2_c \ln\frac{M^2_c}{\mu^2}$\\ [1ex]
\quad(v1) & $0$\\ [1ex]
\quad(w1) & $-\frac{3g^4}{32\pi^4}M^2_c \ln\frac{M^2_c}{\mu^2}$\\ [1ex]
\quad(x1) & $-\frac{g^4}{32\pi^4}M^2_c \ln\frac{M^2_c}{\mu^2}$\\ [1ex]
\quad(y1) & $0$\\ [1ex]
\hline
\\ [-2ex]
\quad Total & $0$\\ [1ex]
\hline
\end{tabular}
\end{table}
%%%%%%%%%%%%%%%%%%%%%%%%%%%%%%%%%%%%%%%%%%%%%%%%%%%%%%%%%%%%%%%%%%%%%%%%%%%%%%%%
The results for the self-energy diagrams of the scalar particle A are given in Table \ref{2LoopAA}. For our current purposes, we only track the quadratic contributions in our calculations.

Here, we see explicitly that the total quadratic contributions vanish. Microscopically, one has
\begin{align}
&(a1)+(c1)+(i1) \sim 0,\\
&(b1)+(d1)+(j1) \sim 0,\\
&(e1)+(f1)+(v1)+(y1) \sim 0,\\
&(g1)+(k1)+(m1)+(o1)+(r1) \sim 0,\\
&(h1)+(l1)+(n1)+(p1)+(q1) \sim 0, \\
&(s1)+(t1)+(u1)+(w1)+(x1) \sim 0,
\end{align}
up to logarithmic contributions and finite terms.

 %%%%%%%%%%%%%%%%%%%%%%%%%%%%%%%%%%%%% FIGURE 3 %%%%%%%%%%%%%%%%%%%%%%%%%%%%%%%%%%%5
\begin{figure}
\centering
\includegraphics[scale=0.9]{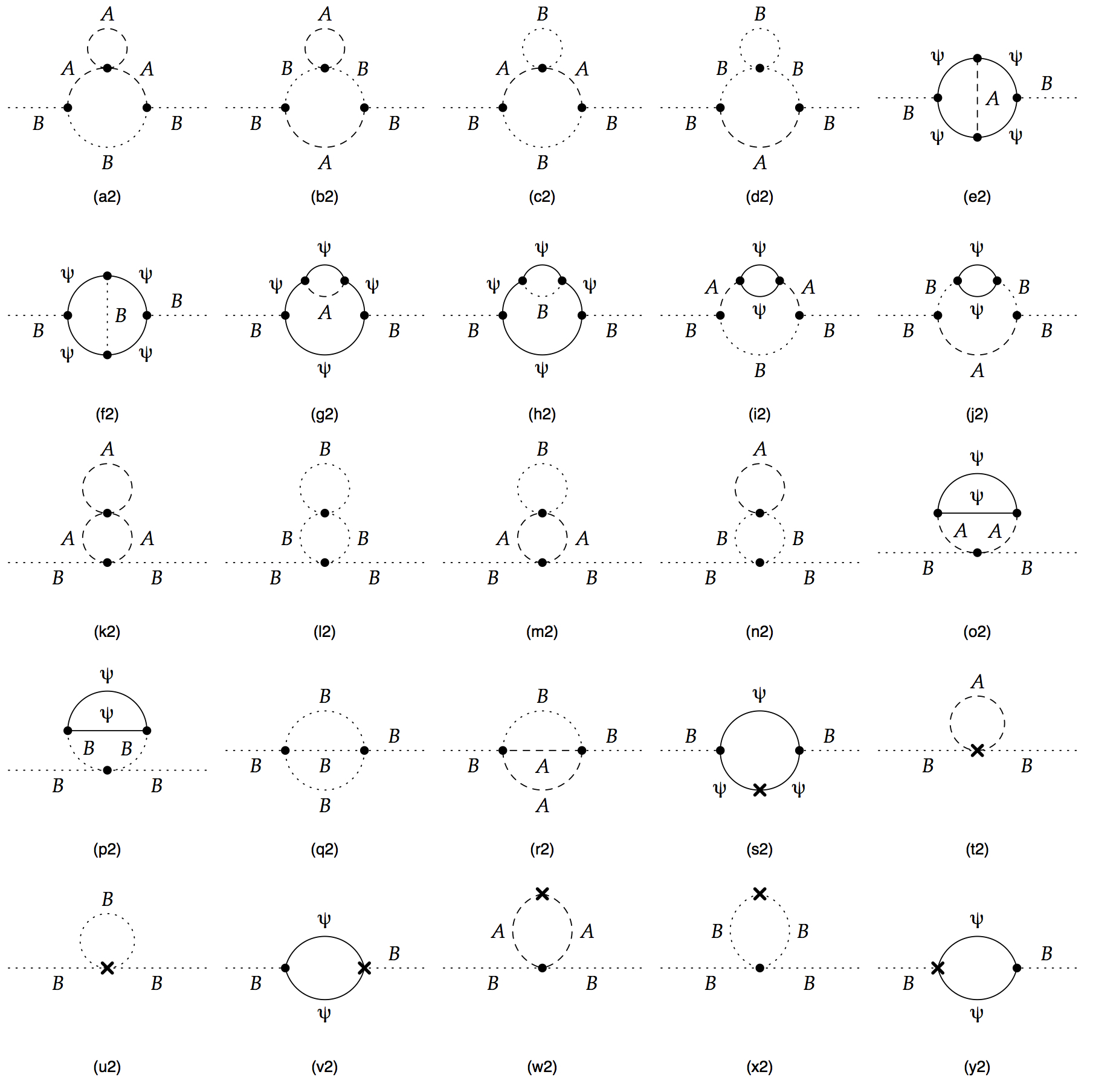}
\caption{Two-loop Self-energy Diagrams for Pseudoscalar Particle B}
\label{2LoopBB}
\end{figure}
%%%%%%%%%%%%%%%%%%%%%%%%%%%%%%%%%%%%%%%%%%%%%%%%%%%%%%%%%%%%%%%%%%%%%%%%%%%%%%%

%%%%%%%%%%%%%%%%%%%%%%%%%%%%%%%%%%%%% TABLE 3 %%%%%%%%%%%%%%%%%%%%%%%%%%%%%%%%%%%%%
\begin{table}
\caption{Quadratic Contributions in the Two-Loop Self-energy Diagrams of Pseudocalar Particle B}
\label{2LoopBB}
\centering
\begin{tabular}{p{5cm} c c}
\\[-2ex]
\hline
\hline
Diagram & Quadratic Contributions\\
\hline
\\[-2ex]
\quad(a2) & $-\frac{3g^2(3\lambda+gm_{\psi})^2}{32\pi^4}M^2_c\int^1_0\!\mathrm{d}x\frac{x}{(1-x)m^2_B+xm^2_A-x(1-x)k_1^2}$\\ [1ex]
\quad(b2) & $-\frac{g^2(3\lambda+gm_{\psi})^2}{32\pi^4}M^2_c\int^1_0\!\mathrm{d}x\frac{x}{(1-x)m^2_A+xm^2_B-x(1-x)k_1^2}$\\ [1ex]
\quad(c2) & $-\frac{g^2(3\lambda+gm_{\psi})^2}{32\pi^4}M^2_c\int^1_0\!\mathrm{d}x\frac{x}{(1-x)m^2_B+xm^2_A-x(1-x)k_1^2}$\\ [1ex]
\quad(d2) & $-\frac{3g^2(3\lambda+gm_{\psi})^2}{32\pi^4}M^2_c\int^1_0\!\mathrm{d}x\frac{x}{(1-x)m^2_A+xm^2_B-x(1-x)k_1^2}$\\ [1ex]
\quad(e2) & $\frac{g^4}{8\pi^4}M^2_c\left[3\left(\ln\frac{M^2_c}{q^2_0}-\gamma_{\omega}\right)+1\right]$\\ [1ex]
\quad(f2) & $-\frac{g^4}{8\pi^4}M^2_c\left[3\left(\ln\frac{M^2_c}{q^2_0}-\gamma_{\omega}\right)+1\right]$\\ [1ex]
\quad(g2) & $-\frac{g^4}{8\pi^4}M^2_c\left[3\left(\ln\frac{M^2_c}{q^2_0}-\gamma_{\omega}\right)+1\right]$\\ [1ex]
\quad(h2) & $-\frac{g^4}{8\pi^4}M^2_c\left[3\left(\ln\frac{M^2_c}{q^2_0}-\gamma_{\omega}\right)+1\right]$\\ [1ex]
\quad(i2) & $\frac{g^2(3\lambda+gm_{\psi})^2}{8\pi^4}M^2_c\int^1_0\!\mathrm{d}x\frac{x}{(1-x)m^2_B+xm^2_A-x(1-x)k_1^2}$\\ [1ex]
\quad(j2) & $\frac{g^2(3\lambda+gm_{\psi})^2}{8\pi^4}M^2_c\int^1_0\!\mathrm{d}x\frac{x}{(1-x)m^2_A+xm^2_B-x(1-x)k_1^2}$\\ [1ex]
\quad(k2) & $\frac{3g^4}{64\pi^4}M^2_c\left(\ln\frac{M^2_c}{m^2_A}-\gamma_{\omega}\right)$\\ [1ex]
\quad(l2) & $\frac{9g^4}{64\pi^4}M^2_c\left(\ln\frac{M^2_c}{m^2_B}-\gamma_{\omega}\right)$\\ [1ex]
\quad(m2) & $\frac{g^4}{64\pi^4}M^2_c\left(\ln\frac{M^2_c}{m^2_A}-\gamma_{\omega}\right)$\\ [1ex]
\quad(n2) & $\frac{3g^4}{64\pi^4}M^2_c\left(\ln\frac{M^2_c}{m^2_B}-\gamma_{\omega}\right)$\\ [1ex]
\quad(o2) & $\frac{g^4}{32\pi^4}M^2_c\left(\ln\frac{M^2_c}{q^2_0}+2\ln\frac{m^2_A}{q^2_0}-\gamma_{\omega}+1\right)$\\ [1ex]
\quad(p2) & $\frac{3g^4}{32\pi^4}M^2_c\left(\ln\frac{M^2_c}{q^2_0}+2\ln\frac{m^2_B}{q^2_0}-\gamma_{\omega}+1\right)$\\ [1ex]
\quad(q2) & $\frac{3g^4}{32\pi^4}M^2_c\left[3\left(\ln\frac{M^2_c}{q^2_0}-\gamma_{\omega}\right)+1\right]$\\ [1ex]
\quad(r2) & $\frac{g^4}{32\pi^4}M^2_c\left[3\left(\ln\frac{M^2_c}{q^2_0}-\gamma_{\omega}\right)+1\right]$\\ [1ex]
\quad(s2) & $\frac{g^4}{4\pi^4}M^2_c\ln\frac{M^2_c}{\mu^2}$\\ [1ex]
\quad(t2) & $-\frac{g^4}{32\pi^4}M^2_c \ln\frac{M^2_c}{\mu^2}$\\ [1ex]
\quad(u2) & $-\frac{3g^4}{32\pi^4}M^2_c \ln\frac{M^2_c}{\mu^2}$\\ [1ex]
\quad(v2) & $0$\\ [1ex]
\quad(w2) & $-\frac{g^4}{32\pi^4}M^2_c\ln\frac{M^2_c}{\mu^2}$\\ [1ex]
\quad(x2) & $-\frac{3g^4}{32\pi^4}M^2_c\ln\frac{M^2_c}{\mu^2}$\\ [1ex]
\quad(y2) & $0$\\ [1ex]
\hline
\\ [-2ex]
\quad Total & $0$\\ [1ex]
\hline
\end{tabular}
\end{table}
%%%%%%%%%%%%%%%%%%%%%%%%%%%%%%%%%%%%%%%%%%%%%%%%%%%%%%%%%%%%%%%%%%%%%%%%%%%%
 
Similar results can be obtained for the pseudoscalar particle B. The corresponding quadratic contributions in the self-energy diagrams of B are given in Table \ref{2LoopBB}.
 
Also, one sees explicitly that the quadratic contributions in self-energy diagrams of B vanish as expected. Microscopically, one has
\begin{align}
&(a2)+(c2)+(i2) \sim 0,\\
&(b2)+(d2)+(j2) \sim 0,\\
&(e2)+(f2) +(v2)+(y2)\sim 0,\\
&(g2)+(k2)+(m2)+(o2)+(r2) \sim 0,\\
&(h2)+(l2)+(n2)+(p2)+(q2) \sim 0, \\
&(s2)+(t2)+(u2)+(w2)+(x2) \sim 0,
\end{align}
up to logarithmic contributions and finite terms.
 
At the two-loop level, Feynman diagrams other than the self-energy diagrams of A and B, although they cannot be overall quadratic divergent by naive power counting, can also contain quadratic subcontributions. These quadratic subcontributions are resulted from the embedding of the one-loop self-energy diagrams of A and B into the Feynman diagrams and thus should cancel with each other since we have shown explicitly that there is no quadratic contribution at the one loop level in Section \ref{1LoopCalculation}. In this way, it is shown by explicit Feynman diagrammatic calculations that up to the two loop level the WZ' model is free of quadratic contributions.

\section{Conclusions and Remarks}
In this article, we revisit the absence of the quadratic contributions in models with softly broken SUSY using LORE. In previous studies, supergraph techniques have been used to show that models with softly broken SUSY should be free of quadratic contributions. Although elegant, supergraph techniques are less useful in phenomenological studies, where the traditional Feynman diagrammatic approach is more suitable. On the other hand, LORE has been shown in previous studies to be a powerful tool to regularize quantum field theories with gauge symmetries and supersymmetry, and is an ideal regularization scheme for traditional Feynman diagrammatic calculations. We use LORE to calculate the two-loop quadratic contributions in WZ' model, the simplest model with softly broken SUSY, which contains a scalar particle A, a pseudoscalar B and a Majorana fermion $\psi$, and show that they do cancel with each other perfectly, which is consistent with the results in literature. Moreover, there should be no obstacle to extend our methods to models containing gauge bosons, such as MSSM, thanks to the fact that LORE preserves gauge symmetries as well. {Also, given the fact that quadratic contributions play a crucial role in deriving the gap equations to describe the dynamically generated spontaneous chiral symmetry breaking of QCD \cite{Dai:2003ip,Gherghetta:1994cr}, the absence of quadratic contributions in SUSY may reveal the fact that its spontaneous breaking has to be carried out in a different manner.} We will return to these issues in future publications. 

\acknowledgments
The author(DB) would like to thank Zhuo Liu for enlightening discussions during the preparation of this work, and G.~Gnendiger and A.~Signer for helpful correspondence after the manuscript was submitted to arXiv. This work was supported in part by the National Science Foundation of China (NSFC) under Grants No.~11690022 \& 11475237, and by the Strategic Priority Research Program of the Chinese Academy of Sciences (CAS), Grant No.~XDB23030100, and by the CAS Center for Excellence in Particle Physics (CCEPP).

%\newpage

\appendix
\section{Useful Formulae for LORE}
\underline{Feynman Parametrization}
\begin{align}
&\frac{1}{AB}=\int^1_0\!\mathrm{d}x\frac{1}{[xA+(1-x)B]^2},\\
&\frac{1}{AB^n}=\int^1_0\!\mathrm{d}x\frac{nx^{n-1}}{[(1-x)A+xB]^{n+1}},
\end{align}
\begin{align}
&\frac{1}{A_1A_2\cdots A_n}=\int^1_0\!\mathrm{d}x_1\cdots\mathrm{d}x_n\delta{(\Sigma x_i-1)}\frac{(n-1)!}{(x_1A_1+x_2A_2+\cdots+x_nA_n)^n},\\
&\frac{1}{A^{m_1}_1A^{m_2}_2\cdots A^{m_n}_n}=\int^1_0\!\mathrm{d}x_1\cdots\mathrm{d}x_n\delta{(\Sigma x_i-1)}\frac{\prod x^{m_i-1}_i}{(\Sigma x_iA_i)^{\Sigma m_i}}\frac{\Gamma(m_1+\cdots+m_n)}{\Gamma(m_1)\cdots\Gamma(m_n)}.
\end{align}
\newline
\newline
\underline{LORE-Regularized ILIs}
\begin{align}
&\int\!\mathrm{d}^4k\frac{1}{k^2-\mathcal{M}^2}\nonumber\\
&:=-i\pi^2\left\{M^2_c-(\mathcal{M}^2+\mu^2_s)\left[\ln\frac{M^2_c}{\mathcal{M}^2+\mu^2_s}-\gamma_{\omega}+1+y_2\left(\frac{\mathcal{M}^2+\mu^2_s}{M^2_c}\right)\right]\right\},\\
&\int\!\mathrm{d}^4k\frac{1}{(k^2-\mathcal{M}^2)^2}=i\pi^2\left[\ln\frac{M^2_c}{\mathcal{M}^2+\mu^2_s}-\gamma_{\omega}+y_0\left(\frac{\mathcal{M}^2+\mu^2_s}{M^2_c}\right)\right],\\
&\int\!\mathrm{d}^4k\frac{1}{(k^2-\mathcal{M}^2)^3}=-i\pi^2\frac{1}{2(\mathcal{M}^2+\mu^2_s)}\left[1-y_{-2}\left(\frac{\mathcal{M}^2+\mu^2_s}{M^2_c}\right)\right],\\
&\int\!\mathrm{d}^4k\frac{1}{(k^2-\mathcal{M}^2)^{\alpha}}\nonumber\\
&=(-1)^{\alpha}i\pi^2\frac{\Gamma(\alpha-2)}{\Gamma(\alpha)}\frac{1}{(\mathcal{M}^2+\mu^2_s)^{\alpha-2}}\left[1-y_{-2(\alpha-2)}\left(\frac{\mathcal{M}^2+\mu^2_s}{M^2_c}\right)\right],
\text{\qquad($\alpha\geq 3$)}\\
&\int\!\mathrm{d}^4k\frac{k^{\mu}k^{\nu}}{(k^2-\mathcal{M}^2)^2}\nonumber\\
&=-\frac{i}{2}g^{\mu\nu}\pi^2\left\{M^2_c-(\mathcal{M}^2+\mu^2_s)\left[\ln\frac{M^2_c}{\mathcal{M}^2+\mu^2_s}-\gamma_{\omega}+1+y_2\left(\frac{\mathcal{M}^2+\mu^2_s}{M^2_c}\right)\right]\right\},\\
&\int\!\mathrm{d}^4k\frac{k^{\mu}k^{\nu}}{(k^2-\mathcal{M}^2)^3}=\frac{i}{4}g^{\mu\nu}\pi^2\left[\ln\frac{M^2_c}{\mathcal{M}^2+\mu^2_s}-\gamma_{\omega}+y_0\left(\frac{\mathcal{M}^2+\mu^2_s}{M^2_c}\right)\right],\\
&\int\!\mathrm{d}^4k\frac{k^{\mu}k^{\nu}}{(k^2-\mathcal{M}^2)^4}=-\frac{i}{6}g^{\mu\nu}\pi^2\frac{1}{2(\mathcal{M}^2+\mu^2_s)}\left[1-y_{-2}\left(\frac{\mathcal{M}^2+\mu^2_s}{M^2_c}\right)\right],\\
&\int\!\mathrm{d}^4k\frac{k^{\mu}k^{\nu}}{(k^2-\mathcal{M}^2)^{\alpha+1}}\nonumber\\
&=(-1)^{\alpha}\frac{i}{2}g^{\mu\nu}\pi^2\frac{\Gamma(\alpha-2)}{\Gamma(\alpha+1)}\frac{1}{(\mathcal{M}^2+\mu^2_s)^{\alpha-2}}\left[1-y_{-2(\alpha-2)}\left(\frac{\mathcal{M}^2+\mu^2_s}{M^2_c}\right)\right],\\
&\int\!\mathrm{d}^4k\frac{k^{\mu}k^{\nu}k^{\rho}k^{\sigma}}{(k^2-\mathcal{M}^2)^3}\nonumber\\
&:=-\frac{i}{8}g^{\{\mu\nu\rho\sigma\}}\pi^2\Bigg\{M^2_c-(\mathcal{M}^2+\mu^2_s)\Bigg[\ln\frac{M^2_c}{\mathcal{M}^2+\mu^2_s}-\gamma_{\omega}+1\nonumber\\
&+y_2\left(\frac{\mathcal{M}^2+\mu^2_s}{M^2_c}\right)\Bigg]\Bigg\},\\
&\int\!\mathrm{d}^4k\frac{k^{\mu}k^{\nu}k^{\rho}k^{\sigma}}{(k^2-\mathcal{M}^2)^4}=\frac{i}{24}g^{\{\mu\nu\rho\sigma\}}\pi^2\left[\ln\frac{M^2_c}{\mathcal{M}^2+\mu^2_s}-\gamma_{\omega}+y_0\left(\frac{\mathcal{M}^2+\mu^2_s}{M^2_c}\right)\right],
\end{align}
\begin{align}
&\int\!\mathrm{d}^4k\frac{k^{\mu}k^{\nu}k^{\rho}k^{\sigma}}{(k^2-\mathcal{M}^2)^5}=-\frac{i}{48}g^{\{\mu\nu\rho\sigma\}}\pi^2\frac{1}{2(\mathcal{M}^2+\mu^2_s)}\left[1-y_{-2}\left(\frac{\mathcal{M}^2+\mu^2_s}{M^2_c}\right)\right],\\
&\int\!\mathrm{d}^4k\frac{k^{\mu}k^{\nu}k^{\rho}k^{\sigma}}{(k^2-\mathcal{M}^{2})^{\alpha+2}}\nonumber\\
&=(-1)^{\alpha}\frac{i}{4}g^{\{\mu\nu\rho\sigma\}}\pi^2\frac{\Gamma(\alpha-2)}{\Gamma(\alpha+2)}\frac{1}{(\mathcal{M}^2+\mu^2_s)^{\alpha-2}}\left[1-y_{-2(\alpha-2)}\left(\frac{\mathcal{M}^2+\mu^2_s}{M^2_c}\right)\right].
\end{align}
Here
\begin{equation}
g^{\{\mu\nu\rho\sigma\}}:=g^{\mu\nu}g^{\rho\sigma}+g^{\mu\rho}g^{\nu\sigma}+g^{\mu\sigma}g^{\rho\nu}.
\end{equation}
$M_c$ and $\mu_s$ in the above formulae regularize the UV contributions and IR divergences respectively. The y functions are given by
\begin{align}
&y_2(x):=\frac{1}{x}(1-x-e^{-x})+\int^x_0\!\mathrm{d}\sigma\frac{1-e^{-\sigma}}{\sigma},\\
&y_0(x)=\int^x_0\!\mathrm{d}\sigma\frac{1-e^{-\sigma}}{\sigma},\\
&y_{-2}(x)=1-e^{-x},\\
&y_{-2(\alpha-1)}(x)=y_{-2(\alpha-2)}(x)-\frac{1}{\alpha-2}x\frac{\partial}{\partial x}y_{-2(\alpha-2)}(x),
\end{align}
for $\alpha \geq 3$. It is easy to show that these y functions approach zero when $x\to 0$.
\newline
\newline
\underline{$\alpha\beta\gamma$ Integrals}
\begin{align}
&I_{121}:=\int\!\mathrm{d}^4l_1\!\int\!\mathrm{d}^4l_2\frac{1}{l^2_1-m^2_1}\frac{1}{(l^2_2-m^2_2)^2}\frac{1}{(l_1+l_2+p)^2-m^2_3}\nonumber\\
&=-\pi^4\left(\ln\frac{M^2_c}{q^2_0}-\gamma_{\omega}+1\right)\left(\ln\frac{M^2_c}{m^2_2}-\gamma_{\omega}\right)+\text{finite terms},
\end{align}
\begin{align}
&I_{111}:=\int\!\mathrm{d}^4l_1\!\int\!\mathrm{d}^4l_2\frac{1}{l^2_1-m^2_1}\frac{1}{l^2_2-m^2_2}\frac{1}{(l_1+l_2+p)^2-m^2_3}\nonumber\\
&=\pi^4M^2_c\left[3\left(\ln\tfrac{M^2_c}{q^2_0}-\gamma_{\omega}\right)+1\right]-\pi^4m^2_2\left(\ln\tfrac{M^2_c}{q^2_0}-\gamma_{\omega}+2\right)\left(\ln\tfrac{M^2_c}{m^2_2}-\gamma_{\omega}+1\right)\nonumber\\
&-\tfrac{\pi^4}{56}[-4\pi^2m^2_1+4\pi^2m^2_3+3(m^2_1-m^2_2)\psi^{(1)}(\tfrac{1}{6})+3(-m^2_2+m^2_3)\psi^{(1)}(\tfrac{1}{3})\nonumber\\
&+3(-m^2_1+m^2_2)\psi^{(1)}(\tfrac{2}{3})+3(m^2_2-m^2_3)\psi^{(1)}(\tfrac{5}{6})]\left(\ln\tfrac{M^2_c}{m^2_2}-\gamma_{\omega}+1\right)\nonumber\\
&-\pi^4m^2_1\left(\ln\tfrac{M^2_c}{m^2_1}-\gamma_{\omega}+1\right)\left(\ln\tfrac{M^2_c}{q^2_0}-\gamma_{\omega}-\tfrac{2\pi}{3\sqrt{3}}\right)\nonumber\\
&-\pi^4m^2_3\left(\ln\tfrac{M^2_c}{m^2_3}-\gamma_{\omega}+1\right)\left(\ln\tfrac{M^2_c}{q^2_0}-\gamma_{\omega}-\tfrac{2\pi}{3\sqrt{3}}\right)\nonumber\\
&-\tfrac{\pi^4}{18}[-54m^2_2+36m^2_3+4\sqrt{3}\pi(m^2_1+m^2_3)\nonumber\\
&+\tfrac{4}{3}\pi^2(m^2_1-m^2_2)+(2m^2_2-m^2_1-m^2_3)\psi^{(1)}(\tfrac{1}{6})+(m^2_2-m^2_3)\psi^{(1)}(\tfrac{1}{3})\nonumber\\
&+(m^1_1-2m^2_2+m^2_3)\psi^{(1)}(\tfrac{2}{3})+(-m^2_2+m^2_3)\psi^{(1)}(\tfrac{5}{6})]\left(\ln\tfrac{M^2_c}{m^2_1(m^2_3)}-\gamma_{\omega}+1\right)\nonumber\\
&+\tfrac{\pi^4}{108}p^2[54(\ln\tfrac{M^2_c}{-p^2}-\gamma_{\omega}+1)+81+2\psi^{(1)}(\tfrac{1}{6})+2\psi^{(1)}(\tfrac{1}{3})-2\psi^{(1)}(\tfrac{2}{3})-2\psi^{(1)}(\tfrac{5}{6})]\nonumber\\
&+\text{finite terms} .
\label{I111}
\end{align}
Here $q^2_0$ is an arbitrary mass scale introduced to balance the dimension, and $\psi^{(1)}(z)=\frac{d^2}{dz^2}\ln\Gamma(z)$ is the first order polygamma function. We have assumed $m^2_1 \sim m^2_3$ in Eq.~\eqref{I111}. It is difficult to obtain an explicit analytic expression of $I_{111}$ for general cases.

\section{Derivations of Eq.~(A.23)}
Here, we give a brief introduction to the UVDP parametrization, using the Feynman integral $I_{111}$ as an explicit example, the only $\alpha\beta\gamma$ integral relevant to the two-loop quadratic contribution calculations. The UVDP methods are aimed to give proper treatments to subcontributions at two loops. In this approach, the UV contributions arising from large loop momenta are transmitted to the asymptotic regions of UVDP parameter space. 
\begin{align}
I_{111}&=\int\!\mathrm{d}^4l_1\int\!\mathrm{d}^4l_2\frac{1}{l_1^2-m_1^2}\frac{1}{l_2^2-m_2^2}\frac{1}{(l_1+l_2+p)^2-m^2_3}\nonumber\\
&=i\pi^2\frac{\Gamma(1)}{\Gamma(1)\Gamma(1)\Gamma(1)}\int_0^\infty\!\prod_{i=1}^3\frac{\mathrm{d}v_i}{(1+v_i)^3}\delta\left(1-\sum_{j=1}^3\frac{1}{1+v_j}\right)\nonumber\\
&\times\frac{(1+v_1)^2(1+v_2)^2(1+v_3)^2}{(3+v_1+v_2+v_3)^2}\int\!\mathrm{d}^4l\frac{1}{l^2-\mathcal{M}^2}\nonumber\\
&=i\pi^2\int_0^\infty\!\prod_{i=1}^3\frac{\mathrm{d}u}{(1+u)^3}\int_0^\infty\!\frac{\mathrm{d}w}{(1+w)^2}\int_0^\infty\!\frac{\mathrm{d}v}{(1+v)^2}\delta\left(1-\frac{1}{1+w}-\frac{1}{1+v}\right)\nonumber\\
&\times\frac{(1+u)^4(1+w)^2(1+v)^2}{\left[u(1+w)(1+v)+1\right]^2}\int\!\mathrm{d}^4l\frac{1}{l^2-\mathcal{M}^2}\nonumber\\
&=\pi^4\int_0^\infty\!\frac{\mathrm{d}w}{(1+w)^2}\int_0^\infty\!\frac{\mathrm{d}v}{(1+v)^2}\delta\left(1-\frac{1}{1+w}-\frac{1}{1+v}\right)\int_0^\infty\!\mathrm{d}u\frac{1+u}{\left[u+\frac{1}{(1+w)(1+v)}\right]^2}\nonumber\\
&\times\left\{M_c^2-M^2\left[\ln\frac{M^2_c}{M^2}-\gamma_\omega+1\right]\right\}\nonumber\\
&=I_{111}^{(0)}+I_{111}^{(1)}+I_{111}^{(2)}+I_{111}^{(3)}+I_{111}^{(4)},
\label{I111AB}
\end{align}
with
\begin{align}
\mathcal{M}^2&=\frac{m_1^2}{1+v_1}+\frac{m^2_2}{1+v_2}+\frac{m^2_3}{1+v_3}-\frac{p^2}{3+v_1+v_2+v_3}\nonumber\\
&=\frac{m_1^2}{(1+u)(1+w)}+\frac{m^2_2u}{1+u}+\frac{m^2_3}{(1+u)(1+v)}-\frac{u}{1+u}\frac{p^2}{u(1+w)(1+v)+1}.
\end{align}
The transition between the old UVDP parameters $(v_1,v_2,v_3)$ and the new one $(u,v,w)$ is given by
\begin{align}
&\frac{1}{1+v_1}=\frac{1}{(1+u)(1+w)},\\
&\frac{1}{1+v_2}=\frac{u}{1+u},\\
&\frac{1}{1+v_3}=\frac{1}{(1+u)(1+v)}.
\end{align}
The following identity plays a crucial role in the UVDP treatment of $I_{111}$ integral,
\begin{align}
\frac{1}{A_1A_2\cdots A_n}=\int_0^\infty\!\prod_{i=1}^n\frac{\mathrm{d}v_i}{(1+v_i)^2}\delta\left(\sum^n_{i=1}\frac{1}{1+v_i}-1\right)\frac{(n-1)!}{\left[\sum_{i=1}^{n}\frac{A_i}{1+v_i}\right]^n}.
\end{align}
The integrals $I_{111}^{(0)}$, $I_{111}^{(1)}$, $I_{111}^{(2)}$, $I_{111}^{(3)}$ and $I_{111}^{(4)}$ introduced in the last step of Eq.~\eqref{I111AB} are analyzed in details as follows. The $I^{(0)}_{111}$ integral contains the quadratic contributions only,
\begin{align}
I_{111}^{(0)}&=\pi^4M_c^2\int_0^\infty\!\frac{\mathrm{d}w}{(1+w)^2}\frac{\mathrm{d}v}{(1+v)^2}\delta\left(1-\frac{1}{1+w}-\frac{1}{1+v}\right)\int_0^\infty\mathrm{d}u\frac{1+u}{\left[u+\frac{1}{(1+w)(1+v)}\right]^2}\nonumber\\
&=\pi^4M_c^2\int_0^\infty\!\frac{\mathrm{d}w}{(1+w)^2}\frac{\mathrm{d}v}{(1+v)^2}\delta\left(1-\frac{1}{1+w}-\frac{1}{1+v}\right)\nonumber\\
&\times\left\{\ln{\frac{M_c^2}{q^2_0}}+\ln(1+w)(1+v)-\gamma_\omega+(1+w)(1+v)-1\right\}\nonumber\\
&=\pi^4M_c^2\int_0^\infty\!\frac{\mathrm{d}w}{(1+w)^2}\left\{\ln\frac{M_c^2}{q^2_0}-\gamma_\omega-1+\ln\left(2+w+\frac{1}{w}\right)+\left(2+\omega+\frac{1}{\omega}\right)\right\}\nonumber\\
&=\pi^4M_c^2\left[3\left(\ln\frac{M^2_c}{q^2_0}-\gamma_\omega\right)+1\right],
\end{align}
while the logarithmic contributions are encapsulated in $I^{(1)}_{111},\cdots,I^{(4)}_{111}$,
\begin{align}
&I^{(1)}_{111}+I^{(2)}_{111}+I^{(3)}_{111}+I^{(4)}_{111}\nonumber\\
&=\pi^4\int_0^\infty\!\frac{\mathrm{d}w}{(1+w)^2}\frac{\mathrm{d}v}{(1+v)^2}\delta\left(1-\frac{1}{1+w}-\frac{1}{1+v}\right)\left\{-\mathcal{M}^2\left[\ln\frac{M_c^2}{\mathcal{M}^2}-\gamma_\omega+1\right]\right\}\nonumber\\
&=-\pi^4\int_0^\infty\!\frac{\mathrm{d}w}{(1+w)^2}\frac{\mathrm{d}v}{(1+v)^2}\delta\left(1-\frac{1}{1+w}-\frac{1}{1+v}\right)\nonumber\\
&\times\int_0^\infty\mathrm{d}u\frac{1+u}{\left[u+\frac{1}{(1+w)(1+v)}\right]^2}\Bigg[\frac{m_1^2}{(1+u)(1+w)}+\frac{m^2_2u}{1+u}+\frac{m_3^2}{(1+u)(1+v)}\nonumber\\
&-\frac{u}{1+u}\frac{p^2}{u(1+w)(1+v)+1}\Bigg]\times\left[\ln\frac{M^2_c}{\mathcal{M}^2}-\gamma_\omega+1\right].
\end{align}
Here, $I_{111}^{(1)},\cdots,I_{111}^{(4)}$ correspond to logarithmic contributions in the following asymptotic UVDP parameter regions:
\begin{itemize}
\item $I^{(1)}_{111}$: $u\to\infty$, $vw=1$, $\mathcal{M}^2\to m^2_2$,
\item $I^{(2)}_{111}$: $v\to\infty$, $u\to0$, $w\to0$, $\mathcal{M}^2\to m^2_1$,
\item $I^{(3)}_{111}$: $w\to\infty$, $u\to0$, $v\to0$, $\mathcal{M}^2\to m^2_3$,
\item $I^{(4)}_{111}$: $-p^2\gg m^2_1, m^2_2, m^2_3$.
\end{itemize}
Explicitly, we have
%\begin{itemize}
%\item
\begin{align}
\bullet\ \,I_{111}^{(1)}&=-\pi^4\int_0^\infty\!\frac{\mathrm{d}w}{(1+w)^2}\frac{\mathrm{d}v}{(1+v)^2}\delta\left(1-\frac{1}{1+w}-\frac{1}{1+v}\right)\nonumber\\
&\times\int_0^\infty\!\mathrm{d}u\frac{1}{u+\frac{1}{(1+w)(1+v)}}\left[\frac{m^2_1}{(1+u)(1+w)}+\frac{m^2_2u}{1+u}+\frac{m^2_3}{(1+u)(1+v)}\right]\nonumber\\
&\times\left[\ln\frac{M^2_c}{m^2_2}-\gamma_\omega+1\right]\nonumber\\
&=-\pi^4m^2_2\left(\ln\frac{M_c^2}{q^2_0}+2-\gamma_\omega\right)\left(\ln\frac{M^2_c}{m^2_2}-\gamma_\omega+1\right)\nonumber
\end{align}
\begin{align}
&-\frac{\pi^4}{56}\Big\{-4m_1^2\pi^2+4m_3^2\pi^2+3(m^2_1-m^2_2)\psi^{(1)}(\tfrac{1}{6})+3(-m_2^2+m^2_3)\psi^{(1)}(\tfrac{1}{3})\nonumber\\
&+3(-m^2_1+m^2_2)\psi^{(1)}(\tfrac{2}{3})+3(m^2_2-m^2_3)\psi^{(1)}(\tfrac{5}{6})\Big\}\left(\ln\frac{M^2_c}{m^2_2}-\gamma_\omega+1\right),
\end{align}
%\item
\begin{align}
\bullet\ \,&I^{(2)}_{111}+I^{(3)}_{111}=-\pi^4\int_0^\infty\!\frac{\mathrm{d}w}{(1+w)^2}\frac{\mathrm{d}v}{(1+v)^2}\delta\left(1-\frac{1}{1+w}-\frac{1}{1+v}\right)\nonumber\\
&\times\int_0^\infty\!\mathrm{d}u\left[1-\frac{1}{(1+w)(1+v)}\right]\frac{1}{\left[u+\frac{1}{(1+w)(1+v)}\right]^2}\bigg[\frac{m^2_1}{(1+w)(1+u)}+\frac{m^2_2u}{1+u}\nonumber\\
&+\frac{m^2_3}{(1+u)(1+v)}\bigg]\left(\ln\frac{M^2_c}{\mathcal{M}^2}-\gamma_\omega+1\right)\nonumber\\
&=-\pi^4m^2_1\left(\ln\frac{M^2_c}{m_1^2}-\gamma_\omega+1\right)\int_0^\infty\!\mathrm{d}v\frac{v^2}{(1+v)(1+v+v^2)}\nonumber\\
&-\pi^4m^2_3\left(\ln\frac{M^2_c}{m_3^2}-\gamma_\omega+1\right)\int_0^\infty\!\mathrm{d}w\frac{w^2}{(1+w)(1+w+w^2)}\nonumber\\
&-\pi^4\int_0^\infty\frac{\mathrm{d}w}{(1+w)^2}\frac{1}{1+w+\frac{1}{w}}\bigg\{(m^2_1-m^2_2+m^2_3)\left(w+\frac{1}{w}\right)+(2m^2_1-m^2_2+2m^2_3)\nonumber\\
&+\left[2m^2_2-m^2_3-m^2_1\left(1+\frac{1}{w}\right)+\frac{m^2_2}{w}\right]\ln\left(w+\frac{1}{w}+2\right)\bigg\}\nonumber\\
&\times\left[\ln M^2_c-\gamma_\omega+1-\ln\left(\frac{m^2_3}{1+\frac{1}{w}}+\frac{m^2_1}{1+w}\right)\right]\nonumber\\
&=-\pi^4m^2_1\left(\ln\frac{M^2_c}{m^2_1}-\gamma_\omega+1\right)\left(\ln\frac{M^2_c}{q^2_0}-\gamma_\omega-\frac{2\pi}{3\sqrt{3}}\right)\nonumber\\
&-\pi^4m^2_3\left(\ln\frac{M^2_c}{m^2_3}-\gamma_\omega+1\right)\left(\ln\frac{M^2_c}{q^2_0}-\gamma_\omega-\frac{2\pi}{3\sqrt{3}}\right)\nonumber\\
&-\frac{\pi^4}{18}\left(\ln\frac{M^2_c}{\mathcal{M}^2}-\gamma_\omega+1\right)\bigg\{-54m^2_2+36m^2_3+4\sqrt{3}(m^2_1+m^2_3)\pi\nonumber\\
&+\frac{4}{3}(m^2_1-m^2_2)\pi^2-(m^2_1-2m^2+m^2_3)\psi^{(1)}(\tfrac{1}{6})+(m^2_2-m^2_3)\psi^{(1)}(\tfrac{1}{3})\nonumber\\
&+(m^2_1-2m^2+m^2_2)\psi^{(1)}(\tfrac{2}{3})+(-m^2_2+m^2_3)\psi^{(1)}(\tfrac{5}{6})\bigg\},
\end{align}
%\item
\begin{align}
\bullet\ \,I_{111}^{(4)}&=\pi^4\int_0^\infty\!\frac{\mathrm{d}w}{(1+w)^2}\frac{\mathrm{d}v}{(1+v)^2}\delta\left(1-\frac{1}{1+w}-\frac{1}{1+v}\right)\nonumber\\
&\times\int_0^\infty\!\mathrm{d}u\frac{u}{(1+w)(1+v)\!\!\left[u+\frac{1}{(1+w)(1+v)}\right]^3}\,p^2\left(\ln\frac{M_c^2}{\mathcal{M}^2}-\gamma_\omega+1\right)\nonumber\\
&=\frac{\pi^4p^2}{108}\bigg\{54\left(\ln\frac{M_c^2}{-p^2}-\gamma_\omega+1\right)+81+\psi^{(1)}(\tfrac{1}{6})+2\psi^{(1)}(\tfrac{1}{3})-2\psi^{(1)}(\tfrac{2}{3})-2\psi^{(1)}(\tfrac{5}{6})\bigg\}.
\end{align}
%\end{itemize}
When putting $I_{111}^{(0)},\cdots,I_{111}^{(4)}$ together, one obtains Eq.~\eqref{I111}.

\section{Details for Feynman Diagrammatic Calculations}
Here, we present some technical details of Feynman diagrammatic calculations by choosing Diagram (e1), (f1), (v1) and (y1) as sample diagrams and calculating their quadratic contributions explicitly. The rest diagrams could be calculated in a similar way.
\begin{align}
(e1)&=-\frac{(-2ig)^4}{512\pi^8}\int\!\mathrm{d}^4q_1\!\int\!\mathrm{d}^4q_2\frac{1}{q_1^2-m_{\psi}^2}\frac{1}{q_2^2-m_{\psi}^2}\frac{1}{(q_1+q_2)^2-m_A^2}\frac{1}{(q_1-k_1)^2-m_{\psi}^2}\nonumber\\
&\times\frac{1}{(q_1+k_1)^2-m_{\psi}^2}\text{tr}\big\{(m_{\psi}+\cancel{q}_2-\cancel{k}_1)(m_\psi+\cancel{q}_2)(m_{\psi}-\cancel{q}_1)(m_\psi-\cancel{q}_1-\cancel{k}_1)\big\}\nonumber\\
&=-\frac{g^4}{8\pi^8}\int_0^1\!\mathrm{d}x\!\int_0^1\!\mathrm{d}y\!\int\!\mathrm{d}^4l_1\!\int\!\mathrm{d}^4l_2\nonumber\\
&\times\frac{1}{\left\{l^2_1+x(1-x)k^2_1-m_{\psi}^2\right\}\left\{l_2^2+y(1-y)k^2_1-m_\psi^2\right\}\left\{[l_1+l_2+(y-x)k_1]^2-m^2_A\right\}}\nonumber\\
&+\text{logarithmic contributions}+\text{finite terms}\nonumber\\
&=-\frac{g^4}{8\pi^8}\int_0^1\!\mathrm{d}x\!\int_0^1\!\mathrm{d}y\pi^4M_c^2\left[3\left(\ln\frac{M^2_c}{q^2_0}-\gamma_\omega\right)+1\right]+\text{logarithmic contributions}\nonumber\\
&+\text{finite terms}\nonumber\\
&=-\frac{g^4}{8\pi^4}M^2_c\left[3\left(\ln\frac{M^2_c}{q^2_0}-\gamma_\omega\right)+1\right]+\text{logarithmic contributions}+\text{finite terms}.
\end{align}
\begin{align}
(f1)&=\frac{(2g)^4}{32\pi^8}\int\!\mathrm{d}^4q_1\!\int\!\mathrm{d}^4q_2\frac{1}{q_1^2-m_\psi^2}\frac{1}{q^2_2-m_\psi^2}\frac{1}{(q_1+q_2)^2-m_B^2}\frac{1}{(q_2-k_1)^2-m_\psi^2}\nonumber\\
&\times\frac{1}{(q_1+k_1)^2-m_\psi^2}\text{tr}\bigg\{(m_\psi+\cancel{q}_2-\cancel{k}_1)(m_\psi+\cancel{q}_2)\gamma^5(m_\psi-\cancel{q}_1)(m_\psi-\cancel{q}_1-\cancel{k}_1)\gamma^5\bigg\}\nonumber\\
&=\frac{g^4}{8\pi^8}\int_0^1\!\mathrm{d}x\!\int_0^1\!\mathrm{d}y\!\int\!\mathrm{d}^4l_1\!\int\!\mathrm{d}^4l_2\nonumber\\
&\times\frac{1}{\left\{l^2_1+x(1-x)k^2_1-m_\psi^2\right\}\left\{l_2^2+y(1-y)k_1^2-m_\psi^2\right\}\left\{[l_1+l_2+(y-x)k_1]^2-m_B^2\right\}}\nonumber\\
&+\text{logarithmic contributions}+\text{finite terms}\nonumber
\end{align}
\begin{align}
&=\frac{g^4}{8\pi^8}\int_0^1\!\mathrm{d}x\!\int_0^1\!\mathrm{d}y\,\pi^4M_c^2\left[3\left(\ln\frac{M_c^2}{q^2_0}-\gamma_\omega\right)+1\right]+\text{logarithmic contributions}\nonumber\\
&+\text{finite terms}\nonumber\\
&=\frac{g^4}{8\pi^4}M^2_c\left[3\left(\ln\frac{M^2_c}{q^2_0}-\gamma_\omega\right)+1\right]+\text{logarithmic contributions}+\text{finite terms}.
\end{align}
The counterterm diagrams (v1) and (y1) are given by
\begin{align}
(v1)=(y1)=0,
\end{align}
as the counterterm vertices $\delta Z^{(1)}_{\bar{\psi}\psi A}=\delta Z^{(1)}_{\bar{\psi}\psi B}=0$. In this article, we are interested in the quadratic contributions only, and have not tracked the logarithmic (sub)contributions and finite terms. The technical subtleties associated with LORE's treatment of overlapping contributions have already been demonstrated in Ref.~\cite{Huang:2011xh,Huang:2012iu}, and we recommend interested readers to go to those references for a comprehensive treatment.
% *********************************************** %
%%%%%%%%%%%%%%%%%%%%%%%%%%%%%%%% END OF APPENDIX %%%%%%%%%%%%%%%%%%%%%%%%%%%%%%
%%%%%%%%%%%%%%%%%%%%%%%%%%%%%%%%%%%%%%%%%%%%%%%%%%%%%%%%%%%%%%%%%%%%%%%%%%%%%
%%%%%%%%%%%%%%%%%%%%%%%%%%%%%%% REFERENCES %%%%%%%%%%%%%%%%%%%%%%%%%%%%%%%%%%%%%%


\begin{thebibliography}{999}


%\cite{Aad:2012tfa}
\bibitem{Aad:2012tfa} 
  G.~Aad {\it et al.} [ATLAS Collaboration],
  ``Observation of a new particle in the search for the Standard Model Higgs boson with the ATLAS detector at the LHC,''
  Phys.\ Lett.\ B {\bf 716}, 1 (2012)
  %doi:10.1016/j.physletb.2012.08.020
  [arXiv:1207.7214 [hep-ex]].
  %%CITATION = doi:10.1016/j.physletb.2012.08.020;%%
  %7262 citations counted in INSPIRE as of 17 May 2017
  
  %\cite{Chatrchyan:2012xdj}
\bibitem{Chatrchyan:2012xdj} 
  S.~Chatrchyan {\it et al.} [CMS Collaboration],
  ``Observation of a new boson at a mass of 125 GeV with the CMS experiment at the LHC,''
  Phys.\ Lett.\ B {\bf 716}, 30 (2012)
  %doi:10.1016/j.physletb.2012.08.021
  [arXiv:1207.7235 [hep-ex]].
  %%CITATION = doi:10.1016/j.physletb.2012.08.021;%%
  %7125 citations counted in INSPIRE as of 17 May 2017

%\cite{Wilson:1973jj}
\bibitem{Wilson:1973jj} 
  K.~G.~Wilson and J.~B.~Kogut,
  ``The Renormalization group and the epsilon expansion,''
  Phys.\ Rept.\  {\bf 12}, 75 (1974).
  %doi:10.1016/0370-1573(74)90023-4
  %%CITATION = doi:10.1016/0370-1573(74)90023-4;%%
  %2372 citations counted in INSPIRE as of 17 May 2017

%\cite{Veltman:1980mj}
\bibitem{Veltman:1980mj} 
  M.~J.~G.~Veltman,
  ``The Infrared - Ultraviolet Connection,''
  Acta Phys.\ Polon.\ B {\bf 12}, 437 (1981).
  %%CITATION = APPOA,B12,437;%%
  %637 citations counted in INSPIRE as of 17 May 2017

%\cite{Bai:2014lea}
\bibitem{Bai:2014lea} 
  D.~Bai, J.~W.~Cui and Y.~L.~Wu,
  ``Quantum Electroweak Symmetry Breaking Through Loop Quadratic Contributions,''
  Phys.\ Lett.\ B {\bf 746}, 379 (2015)
  %doi:10.1016/j.physletb.2015.05.037
  [arXiv:1412.3562 [hep-ph]].
  %%CITATION = doi:10.1016/j.physletb.2015.05.037;%%
  %4 citations counted in INSPIRE as of 17 May 2017
  
  %\cite{Olive:2016xmw}
\bibitem{Olive:2016xmw} 
  C.~Patrignani {\it et al.} [Particle Data Group],
  ``Review of Particle Physics,''
  Chin.\ Phys.\ C {\bf 40}, no. 10, 100001 (2016).
  %doi:10.1088/1674-1137/40/10/100001
  %%CITATION = doi:10.1088/1674-1137/40/10/100001;%%
  %849 citations counted in INSPIRE as of 17 May 2017

  %\cite{Girardello:1981wz}
\bibitem{Girardello:1981wz} 
  L.~Girardello and M.~T.~Grisaru,
  ``Soft Breaking of Supersymmetry,''
  Nucl.\ Phys.\ B {\bf 194}, 65 (1982).
  %doi:10.1016/0550-3213(82)90512-0
  %%CITATION = doi:10.1016/0550-3213(82)90512-0;%%
  %864 citations counted in INSPIRE as of 17 May 2017
  
  %\cite{tHooft:1972tcz}
\bibitem{tHooft:1972tcz} 
  G.~'t Hooft and M.~J.~G.~Veltman,
  ``Regularization and Renormalization of Gauge Fields,''
  Nucl.\ Phys.\ B {\bf 44}, 189 (1972).
  %doi:10.1016/0550-3213(72)90279-9
  %%CITATION = doi:10.1016/0550-3213(72)90279-9;%%
  %3797 citations counted in INSPIRE as of 17 May 2017

%\cite{Siegel:1979wq}
\bibitem{Siegel:1979wq} 
  W.~Siegel,
  ``Supersymmetric Dimensional Regularization via Dimensional Reduction,''
  Phys.\ Lett.\  {\bf 84B}, 193 (1979).
  %doi:10.1016/0370-2693(79)90282-X
  %%CITATION = doi:10.1016/0370-2693(79)90282-X;%%
  %935 citations counted in INSPIRE as of 17 May 2017

  
    %\cite{Martin:1993yx}
\bibitem{Martin:1993yx} 
  S.~P.~Martin and M.~T.~Vaughn,
  ``Regularization dependence of running couplings in softly broken supersymmetry,''
  Phys.\ Lett.\ B {\bf 318}, 331 (1993)
  %doi:10.1016/0370-2693(93)90136-6
  [hep-ph/9308222].
  %%CITATION = doi:10.1016/0370-2693(93)90136-6;%%
  %264 citations counted in INSPIRE as of 30 May 2017
  
  %\cite{Wu:2002xa}
\bibitem{Wu:2002xa} 
  Y.~L.~Wu,
  ``Symmetry principle preserving and infinity free regularization and renormalization of quantum field theories and the mass gap,''
  Int.\ J.\ Mod.\ Phys.\ A {\bf 18}, 5363 (2003)
  %doi:10.1142/S0217751X03015222
  [hep-th/0209021].
  %%CITATION = doi:10.1142/S0217751X03015222;%%
  %35 citations counted in INSPIRE as of 17 May 2017
  
    %\cite{Wu:2003dd}
\bibitem{Wu:2003dd} 
  Y.~L.~Wu,
  ``Symmetry preserving loop regularization and renormalization of QFTs,''
  Mod.\ Phys.\ Lett.\ A {\bf 19}, 2191 (2004)
  %doi:10.1142/S0217732304015361
  [hep-th/0311082].
  %%CITATION = doi:10.1142/S0217732304015361;%%
  %30 citations counted in INSPIRE as of 17 May 2017
    
  %\cite{Cui:2008uv}
\bibitem{Cui:2008uv} 
  J.~W.~Cui and Y.~L.~Wu,
  ``One-Loop Renormalization of Non-Abelian Gauge Theory and beta Function Based on Loop Regularization Method,''
  Int.\ J.\ Mod.\ Phys.\ A {\bf 23}, 2861 (2008)
  %doi:10.1142/S0217751X08040305
  [arXiv:0801.2199 [hep-ph]].
  %%CITATION = doi:10.1142/S0217751X08040305;%%
  %11 citations counted in INSPIRE as of 17 May 2017
  
    %\cite{Dai:2003ip}
\bibitem{Dai:2003ip} 
  Y.~B.~Dai and Y.~L.~Wu,
  ``Dynamically spontaneous symmetry breaking and masses of lightest nonet scalar mesons as composite Higgs bosons,''
  Eur.\ Phys.\ J.\ C {\bf 39}, S1 (2005)
  %doi:10.1140/epjcd/s2004-01-001-3
  [hep-ph/0304075].
  %%CITATION = doi:10.1140/epjcd/s2004-01-001-3;%%
  %45 citations counted in INSPIRE as of 17 May 2017

  %\cite{Tang:2008ah}
\bibitem{Tang:2008ah} 
  Y.~Tang and Y.~L.~Wu,
  ``Gravitational Contributions to the Running of Gauge Couplings,''
  Commun.\ Theor.\ Phys.\  {\bf 54}, 1040 (2010)
  %doi:10.1088/0253-6102/54/6/15
  [arXiv:0807.0331 [hep-ph]].
  %%CITATION = doi:10.1088/0253-6102/54/6/15;%%
  %32 citations counted in INSPIRE as of 17 May 2017
  
    %\cite{Tang:2010cr}
\bibitem{Tang:2010cr} 
  Y.~Tang and Y.~L.~Wu,
  ``Quantum Gravitational Contributions to Gauge Field Theories,''
  Commun.\ Theor.\ Phys.\  {\bf 57}, 629 (2012)
  %doi:10.1088/0253-6102/57/4/19
  [arXiv:1012.0626 [hep-ph]].
  %%CITATION = doi:10.1088/0253-6102/57/4/19;%%
  %13 citations counted in INSPIRE as of 17 May 2017
  
    %\cite{Tang:2011gz}
\bibitem{Tang:2011gz} 
  Y.~Tang and Y.~L.~Wu,
  ``Gravitational Contributions to Gauge Green's Functions and Asymptotic Free Power-Law Running of Gauge Coupling,''
  JHEP {\bf 1111}, 073 (2011)
  %doi:10.1007/JHEP11(2011)073
  [arXiv:1109.4001 [hep-ph]].
  %%CITATION = doi:10.1007/JHEP11(2011)073;%%
  %8 citations counted in INSPIRE as of 17 May 2017

  
  %\cite{Cui:2008bk}
\bibitem{Cui:2008bk} 
  J.~W.~Cui, Y.~Tang and Y.~L.~Wu,
  ``Renormalization of Supersymmetric Field Theories in Loop Regularization with String-mode Regulators,''
  Phys.\ Rev.\ D {\bf 79}, 125008 (2009)
  %doi:10.1103/PhysRevD.79.125008
  [arXiv:0812.0892 [hep-ph]].
  %%CITATION = doi:10.1103/PhysRevD.79.125008;%%
  %8 citations counted in INSPIRE as of 17 May 2017
    
  %\cite{Cui:2011za}
\bibitem{Cui:2011za} 
  J.~W.~Cui, Y.~L.~Ma and Y.~L.~Wu,
  ``Explicit derivation of the QED trace anomaly in symmetry-preserving loop regularization at one-loop level,''
  Phys.\ Rev.\ D {\bf 84}, 025020 (2011)
  %doi:10.1103/PhysRevD.84.025020
  [arXiv:1103.2026 [hep-ph]].
  %%CITATION = doi:10.1103/PhysRevD.84.025020;%%
  %6 citations counted in INSPIRE as of 17 May 2017
  
  %\cite{Huang:2011yf}
\bibitem{Huang:2011yf} 
  D.~Huang, Y.~Tang and Y.~L.~Wu,
  ``Note on Higgs Decay into Two Photons $H\to \gamma\gamma$,''
  Commun.\ Theor.\ Phys.\  {\bf 57}, 427 (2012)
  %doi:10.1088/0253-6102/57/3/14
  [arXiv:1109.4846 [hep-ph]].
  %%CITATION = doi:10.1088/0253-6102/57/3/14;%%
  %23 citations counted in INSPIRE as of 17 May 2017
  
  %\cite{Huang:2011xh}
\bibitem{Huang:2011xh} 
  D.~Huang and Y.~L.~Wu,
  ``Consistency and Advantage of Loop Regularization Method Merging with Bjorken-Drell's Analogy Between Feynman Diagrams and Electrical Circuits,''
  Eur.\ Phys.\ J.\ C {\bf 72}, 2066 (2012)
  %doi:10.1140/epjc/s10052-012-2066-2
  [arXiv:1108.3603 [hep-ph]].
  %%CITATION = doi:10.1140/epjc/s10052-012-2066-2;%%
  %6 citations counted in INSPIRE as of 17 May 2017
  
    %\cite{Huang:2012iu}
\bibitem{Huang:2012iu} 
  D.~Huang, L.~F.~Li and Y.~L.~Wu,
  ``Consistency of Loop Regularization Method and Divergence Structure of QFTs Beyond One-Loop Order,''
  Eur.\ Phys.\ J.\ C {\bf 73}, no. 4, 2353 (2013)
  %doi:10.1140/epjc/s10052-013-2353-6
  [arXiv:1210.2794 [hep-ph]].
  %%CITATION = doi:10.1140/epjc/s10052-013-2353-6;%%
  %3 citations counted in INSPIRE as of 17 May 2017
  
  %\cite{Wu:2013iga}
\bibitem{Wu:2013iga} 
  Y.~L.~Wu,
  ``Quantum Structure of Field Theory and Standard Model Based on Infinity-free Loop Regularization/Renormalization,''
  Int.\ J.\ Mod.\ Phys.\ A {\bf 29}, 1430007 (2014)
  %doi:10.1142/9789814590112_0006, 10.1142/S0217751X14300075
  [arXiv:1312.1403 [hep-th]].
  %%CITATION = doi:10.1142/9789814590112_0006, 10.1142/S0217751X14300075;%%
  %4 citations counted in INSPIRE as of 17 May 2017
  
  %\cite{Chapling:2016kpi}
\bibitem{Chapling:2016kpi} 
  R.~Chapling,
  ``Asymptotics of Certain Sums Required in Loop Regularisation,''
  Mod.\ Phys.\ Lett.\ A {\bf 31}, no. 04, 1650030 (2016)
  %doi:10.1142/S0217732316500309
  [arXiv:1601.04966 [hep-th]].
  %%CITATION = doi:10.1142/S0217732316500309;%%
  %1 citations counted in INSPIRE as of 17 May 2017
  
  %\cite{Chapling:2016sfj}
\bibitem{Chapling:2016sfj} 
  R.~Chapling,
  ``Note on Closed-Form Expressions for Irreducible Loop Integrals,''
  arXiv:1608.05311 [hep-th].
  %%CITATION = ARXIV:1608.05311;%%
  
    %\cite{Wu:2017xzu}
\bibitem{Wu:2017xzu} 
  Y.~L.~Wu,
  ``Maximal symmetry and mass generation of Dirac fermion and gravitational gauge field theory in six-dimensional spacetime,''
  arXiv:1703.05436 [hep-th].
  %%CITATION = ARXIV:1703.05436;%%
  
  %\cite{Wu:2017rmd}
\bibitem{Wu:2017rmd} 
  Y.~L.~Wu,
  ``Unified field theory of basic forces and elementary particles with gravitational origin of gauge symmetry in hyper-spacetime,''
  arXiv:1705.06365 [hep-th].
  %%CITATION = ARXIV:1705.06365;%%

%\cite{Pauli:1949zm}
\bibitem{Pauli:1949zm} 
  W.~Pauli and F.~Villars,
  ``On the Invariant regularization in relativistic quantum theory,''
  Rev.\ Mod.\ Phys.\  {\bf 21}, 434 (1949).
  %doi:10.1103/RevModPhys.21.434
  %%CITATION = doi:10.1103/RevModPhys.21.434;%%
  %501 citations counted in INSPIRE as of 17 May 2017

%\cite{Iliopoulos:1974zv}
\bibitem{Iliopoulos:1974zv} 
  J.~Iliopoulos and B.~Zumino,
  ``Broken Supergauge Symmetry and Renormalization,''
  Nucl.\ Phys.\ B {\bf 76}, 310 (1974).
  %doi:10.1016/0550-3213(74)90388-5
  %%CITATION = doi:10.1016/0550-3213(74)90388-5;%%
  %621 citations counted in INSPIRE as of 17 May 2017
  
    %\cite{Gnendiger:2017pys}
\bibitem{Gnendiger:2017pys} 
  C.~Gnendiger {\it et al.},
  ``To ${d}$, or not to ${d}$: recent developments and comparisons of regularization schemes,''
  Eur.\ Phys.\ J.\ C {\bf 77}, no. 7, 471 (2017)
  %doi:10.1140/epjc/s10052-017-5023-2
  [arXiv:1705.01827 [hep-ph]].
  %%CITATION = doi:10.1140/epjc/s10052-017-5023-2;%%
  %3 citations counted in INSPIRE as of 11 Aug 2017
  
  %\cite{Battistel:1998sz}
\bibitem{Battistel:1998sz} 
  O.~A.~Battistel, A.~L.~Mota and M.~C.~Nemes,
  ``Consistency conditions for 4-D regularizations,''
  Mod.\ Phys.\ Lett.\ A {\bf 13}, 1597 (1998).
  %doi:10.1142/S0217732398001686
  %%CITATION = doi:10.1142/S0217732398001686;%%
  %44 citations counted in INSPIRE as of 11 Aug 2017
  
  %\cite{BaetaScarpelli:2001ix}
\bibitem{BaetaScarpelli:2001ix} 
  A.~P.~Baeta Scarpelli, M.~Sampaio, B.~Hiller and M.~C.~Nemes,
  ``Chiral anomaly and CPT invariance in an implicit momentum space regularization framework,''
  Phys.\ Rev.\ D {\bf 64}, 046013 (2001)
  %doi:10.1103/PhysRevD.64.046013
  [hep-th/0102108].
  %%CITATION = doi:10.1103/PhysRevD.64.046013;%%
  %74 citations counted in INSPIRE as of 11 Aug 2017
  
  %\cite{BaetaScarpelli:2000zs}
\bibitem{BaetaScarpelli:2000zs} 
  A.~P.~Baeta Scarpelli, M.~Sampaio and M.~C.~Nemes,
  ``Consistency relations for an implicit n-dimensional regularization scheme,''
  Phys.\ Rev.\ D {\bf 63}, 046004 (2001)
  %doi:10.1103/PhysRevD.63.046004
  [hep-th/0010285].
  %%CITATION = doi:10.1103/PhysRevD.63.046004;%%
  %45 citations counted in INSPIRE as of 11 Aug 2017

%\cite{Pittau:2012zd}
\bibitem{Pittau:2012zd} 
  R.~Pittau,
  ``A four-dimensional approach to quantum field theories,''
  JHEP {\bf 1211}, 151 (2012)
  %doi:10.1007/JHEP11(2012)151
  [arXiv:1208.5457 [hep-ph]].
  %%CITATION = doi:10.1007/JHEP11(2012)151;%%
  %24 citations counted in INSPIRE as of 11 Aug 2017

%\cite{Siegel:1980qs}
\bibitem{Siegel:1980qs} 
  W.~Siegel,
  ``Inconsistency of Supersymmetric Dimensional Regularization,''
  Phys.\ Lett.\  {\bf 94B}, 37 (1980).
  %doi:10.1016/0370-2693(80)90819-9
  %%CITATION = doi:10.1016/0370-2693(80)90819-9;%%
  %320 citations counted in INSPIRE as of 17 May 2017
  
    %\cite{Bjorken:1965zz}
\bibitem{Bjorken:1965zz} 
  J.~D.~Bjorken and S.~D.~Drell,
  ``Relativistic quantum fields.''
  %%CITATION = INSPIRE-873163;%%
  %56 citations counted in INSPIRE as of 30 May 2017


%\cite{Wess:1974tw}
\bibitem{Wess:1974tw} 
  J.~Wess and B.~Zumino,
  ``Supergauge Transformations in Four-Dimensions,''
  Nucl.\ Phys.\ B {\bf 70}, 39 (1974).
  %doi:10.1016/0550-3213(74)90355-1
  %%CITATION = doi:10.1016/0550-3213(74)90355-1;%%
  %2637 citations counted in INSPIRE as of 17 May 2017
  
  %\cite{Wess:1973kz}
\bibitem{Wess:1973kz} 
  J.~Wess and B.~Zumino,
  ``A Lagrangian Model Invariant Under Supergauge Transformations,''
  Phys.\ Lett.\  {\bf 49B}, 52 (1974).
  %doi:10.1016/0370-2693(74)90578-4
  %%CITATION = doi:10.1016/0370-2693(74)90578-4;%%
  %1424 citations counted in INSPIRE as of 17 May 2017
  
  %\cite{Peskin:1995ev}
\bibitem{Peskin:1995ev} 
  M.~E.~Peskin and D.~V.~Schroeder,
  ``An Introduction to quantum field theory,''
  Reading, USA: Addison-Wesley (1995) 842 p
  %1028 citations counted in INSPIRE as of 17 May 2017

%\cite{Denner:1992vza}
\bibitem{Denner:1992vza} 
  A.~Denner, H.~Eck, O.~Hahn and J.~Kublbeck,
  ``Feynman rules for fermion number violating interactions,''
  Nucl.\ Phys.\ B {\bf 387}, 467 (1992).
  %doi:10.1016/0550-3213(92)90169-C
  %%CITATION = doi:10.1016/0550-3213(92)90169-C;%%
  %200 citations counted in INSPIRE as of 17 May 2017

%\cite{Denner:1992me}
\bibitem{Denner:1992me} 
  A.~Denner, H.~Eck, O.~Hahn and J.~Kublbeck,
  ``Compact Feynman rules for Majorana fermions,''
  Phys.\ Lett.\ B {\bf 291}, 278 (1992).
  %doi:10.1016/0370-2693(92)91045-B
  %%CITATION = doi:10.1016/0370-2693(92)91045-B;%%
  %77 citations counted in INSPIRE as of 17 May 2017

%\cite{Hahn:2000kx}
\bibitem{Hahn:2000kx} 
  T.~Hahn,
  ``Generating Feynman diagrams and amplitudes with FeynArts 3,''
  Comput.\ Phys.\ Commun.\  {\bf 140}, 418 (2001)
  %doi:10.1016/S0010-4655(01)00290-9
  [hep-ph/0012260].
  %%CITATION = doi:10.1016/S0010-4655(01)00290-9;%%
  %1193 citations counted in INSPIRE as of 17 May 2017
  
  %\cite{Gherghetta:1994cr}
\bibitem{Gherghetta:1994cr} 
  T.~Gherghetta,
  ``Regularization in the gauged Nambu-Jona-Lasinio model,''
  Phys.\ Rev.\ D {\bf 50}, 5985 (1994)
  %doi:10.1103/PhysRevD.50.5985
  [hep-ph/9408225].
  %%CITATION = doi:10.1103/PhysRevD.50.5985;%%
  %28 citations counted in INSPIRE as of 30 May 2017




%
%% ****** End of file articletemplate.tex ****** %%

\end{thebibliography}
\end{document}